\documentclass[pra,showpacs,preprintnumbers,amsmath,amssymb,tightenlines,twocolumn,superscriptaddress]{revtex4-1}
\usepackage{graphicx}% Include figure files
\usepackage{dcolumn}% Align table columns on decimal point
\usepackage{bm}% bold math
\usepackage{epsfig}
\usepackage{amsmath}
\newcommand{\bra}[1]{\langle\,{#1}\, |}
\newcommand{\ket}[1]{|\,{#1}\,\rangle}

\newcommand{\sub}[2]{{#1}_{\mbox{\!\! \scriptsize #2}}}

\def\beq{\begin{equation}}
	\def\eeq{\end{equation}}

\newcommand{\rref}[1]{ref.~\cite{#1}}
\newcommand{\fref}[1]{Fig.~\ref{#1}}
\newcommand{\frefp}[2]{Fig.~\ref{#1}~(#2)}

\newcommand{\eref}[1]{Eq.~(\ref{#1})}

\newcommand{\sref}[1]{section~\ref{#1}}

\newcommand{\cref}[1]{chapter~\ref{#1}}

\newcommand{\Cref}[1]{Chapter~\ref{#1}}

\newcommand{\aref}[1]{appendix~\ref{#1}}
\newcommand{\bref}[1]{(\ref{#1})}

\newcommand{\beginsupplement}{%
	\setcounter{table}{0}
	\renewcommand{\thetable}{A\arabic{table}}%
	\setcounter{figure}{0}
	\renewcommand{\thefigure}{A\arabic{figure}}%
	\setcounter{equation}{0}
	\def\theequation{A\arabic{equation}}
}

\usepackage{ulem}  %unter- (\uline{important}) und durchstreichen (\sout{wrong})
\usepackage{color}

\begin{document}
\title{Quantum network tomography of Rydberg arrays by machine learning}
\author{K.~Mukherjee}
\affiliation{Department of Physics, Indian Institute of Science Education and Research, Bhopal, Madhya Pradesh 462 066, India}
\affiliation{Homer L. Dodge Department of Physics and Astronomy, The University of Oklahoma, Norman, Oklahoma 73019, USA}
\affiliation{Center for Quantum Research and Technology, The University of Oklahoma, Norman, Oklahoma 73019, USA}
\author{J.~Schachenmayer}
\affiliation{CESQ/ISIS (UMR 7006), University of Strasbourg and CNRS, 67000 Strasbourg, France}
\author{S.~Whitlock}
\affiliation{ISIS (UMR 7006), University of Strasbourg and CNRS, 67000 Strasbourg, France}
\author{S.~W\"uster}
\affiliation{Department of Physics, Indian Institute of Science Education and Research, Bhopal, Madhya Pradesh 462 066, India}
\affiliation{PlanQC GmbH, Lichtenbergstr. 8, 85748 Garching, Germany}
\email{sebastian@iiserb.ac.in}
\begin{abstract}
Configurable arrays of optically trapped Rydberg atoms are a versatile platform for quantum computation and quantum simulation, also allowing controllable decoherence.
We demonstrate theoretically, that they also enable proof-of-principle demonstrations for a technique to build models for open quantum dynamics by machine learning with artificial neural networks, recently proposed in [Mukherjee {\it et al.}~https://arxiv.org/abs/2409.18822 (2024)]. Using the outcome of quantum transport through a network of sites that correspond to excited Rydberg atoms, the multi-stage neural network algorithm successfully identifies the number of atoms (or nodes in the network), and subsequently their location. It further extracts an effective interaction Hamiltonian and decoherence operators induced by the environment.
To probe the Rydberg array, one initiates dynamics repeatedly from the same initial state and then measures the transport probability to 
an output atom. Large datasets are generated by varying the position of the latter. Measurements are required in only one single basis, making the approach complementary to e.g.~quantum process tomography. The cold atom platform discussed in this article can be used to explore the performance of the proposed protocol when training the neural network with simulation data, but then applying it to construct models based on experimental data.
\end{abstract}

\maketitle

\section{Introduction}
\label{NN_NN_NN_intro}
%%%%%%%%%%%%%%%%%%%%%%%%%%%%%%%%%%%
%%%%%%%%%%%%%%%%%%%%%%%%%%%%%%%%%%%
%
In recent years, machine learning (ML) has become a popular and versatile tool in a wide range of scientific disciplines, from natural language processing \cite{hoy2018alexa} over quantum chemistry \cite{raucci2021voice}, bio-chemistry \cite{wei2019protein}, cancer diagnosis \cite{mccarthy2004applications,agarap2018breast,saba2020recent}, developmental biology \cite{bhavna:NNridges} to quantum information theory \cite{wittek2014quantum,banchi2021generalization}. Conversely, the development of neural networks has been boosted by the requirements of quantum technology and quantum chemistry \cite{ceriotti2021introduction}, where neural network algorithms have been used to produce accurate and effective results. Presently, they are increasingly employed to investigate the features of complex quantum systems \cite{hase2017machine,torlai2019integrating,chong2022machine,papivc2022neural,luo2022autoregressive} and superconducting qubits \cite{bandyopadhyay2018applications}.

In a companion article \cite{mukherjee:modelbuilding}, we have further broadened this scope by introducing a scheme through which ML can deduce physical models for open quantum systems from a restricted set of measurements of quantum dynamics traversing a system of interest. This goes beyond quantum process tomography \cite{Poyatos_QPT,Chuang_Nielsen_QPT}, by not just determining a map between input and output Hilbertspaces, but inferring all physical parameters controlling dynamics in-between, and requiring much less diverse measurements. 

\begin{figure}[htb]
	\centering
	\includegraphics[width=0.99\linewidth]{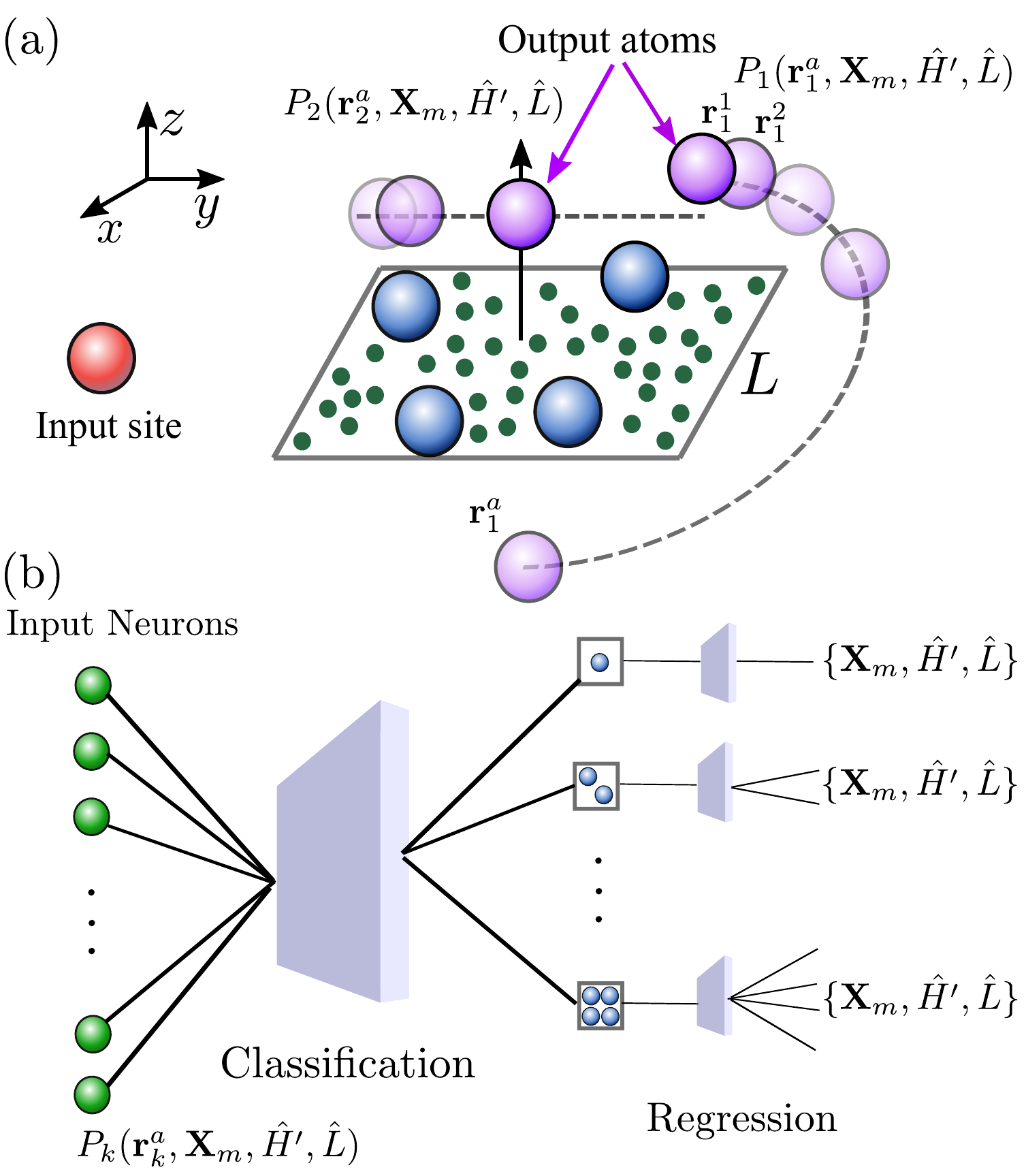}
	\caption{Schematic of setup and multi-branch pipeline for Rydberg quantum network tomography. (a) An assembly of Rydberg atoms (large spheres), with a single Rydberg $p$-state at the \emph{input site} (red sphere).
	The blue spheres inside the box region of side-length $L$ show atoms initially in a Rydberg $s$-state. These atoms define the quantum network to be probed, with small green circles representing background gas atoms that provide controllable decoherence. Violet spheres are the \emph{output} Rydberg $s$-state atoms, the positions of which can be varied along dashed lines as sketched. Measurements of the transport probabilities $P_{1,2}$ from input to output sites for the range of positions then generate large data-sets. (b) These provide the input to a machine learning based multi-branch pipeline, which first classifies the system in the box based on the number of its network nodes $M$ and then uses multi-target regression for each class to locate the position of the atoms ($\mathbf{X}_m$) and matrix elements of key operators such as effective Hamiltonian $\hat{H}'$ and Lindblad decay operator $\hat{L}$ based on the excitation transport probabilities $P_k$. The latter steps all leverage an artificial neural network (ANN).
} 
	\label{NN_models}
\end{figure}
While \rref{mukherjee:modelbuilding} demonstrates the general utility of the scheme without focussing on a specific physical platform, here we apply ML-based-model-building to networks of interacting Rydberg atoms in the presence of a controllable cold gas environment \cite{David:Rydagg,schempp:spintransport}, focussing on realistic parameters. We thus show that Rydberg excitation in optical tweezer arrays
\cite{nogrette2014single,barredo2016atom,endres2016atom,barredo2018synthetic,Wang_Rydberg_array_NPJ}
within an ultra-cold thermal gas allows experiments at the present state-of-the-art to explore key facets of ML based model building: 
utilizing neural networks that have been trained on a large set of computer simulations of an experiment to propose models of the experiment.
Since good models describing the Rydberg experimental platform are actually known, this can then verify the feasibility of the approach. 
After verification of the concept, it can then be further applied to complex quantum devices or systems where a few state effective model should exist but is not a priori known.

Following \rref{mukherjee:modelbuilding}, we employ a multi-branch pipeline capable of probing a quantum system based on its excitation transport properties.
The main branch classifies the system based on the number of Rydberg atoms ($M$) in the network, resulting in $M$ sub-branches, where the coordinates of those Rydberg atoms and the matrix elements of Hamiltonian and Lindbladians are determined.  Together, this reconstructs the entire quantum network and its properties, thus implementing quantum network tomography on the system \cite{de2022quantum}. We then explore the performance of this approach for varying systems sizes and decoherence strengths.
Deterministic positioning of Rydberg atoms using optical tweezers \cite{nogrette2014single,barredo2016atom,endres2016atom,barredo2018synthetic,Wang_Rydberg_array_NPJ}, combined with tunability of long-range dipolar interactions and site-resolved measurements, Rydberg arrays are an ideal platform for the testing of machine learning algorithms designed to analyse quantum systems and their dynamics. The validated approach can then be extended and generalized to arbitrary quantum networks.

This article is organized as follows: in \sref{sec:NN_system}, we introduce the quantum network formed with Rydberg atoms interacting via dipole-dipole interactions. The network is divided in two parts: a ``black-box" which will be taken as inaccessible to measurement but for which a model is sought, and a second part involving atoms that are accessible to measurements and initialisation. In \sref{NN_classification_regression}, we present a multi-branch pipeline capable of probing the inaccessible quantum system based on its excitation transport properties. The root element, discussed in \sref{sec:NN_class_results}, classifies the system based on the number of Rydberg atoms ($M$) and spawns $M$ sub-branches. Then, we discuss in \sref{sec_NN:regression_QNR} how the branches find the coordinates of Rydberg atoms, and in \sref{sec:NN_Heff_Leff} how they infer the matrix elements of Hamiltonian and Lindbladians. We conclude in \sref{concl}.

%%%%%%%%%%%%%%%%%%%%%%%%%%%%%%%%%%%
\section{Rydberg quantum networks}
\label{sec:NN_system}
%%%%%%%%%%%%%%%%%%%%%%%%%%%%%%%%%%%

We consider an assembly of $N$ Rydberg atoms, which we term Rydberg aggregate due to collective excited states \cite{wuester:review}. The atoms are assumed trapped and immobile at controllable locations, for example using configurable optical tweezer arrays \cite{zhang2011magic,Wang_Rydberg_array_NPJ}. 
Each atom can be in two internal states: $\ket{s}\equiv\ket{\nu,l=0}$ and $\ket{p}\equiv\ket{\nu,l=1}$, where $l$ is the angular momentum quantum number. We choose a principal quantum number of $\nu=43$, in order to utilise interaction strength calculations from \rref{David:Rydagg}.
Within the aggregate of $N=M+3$ Rydberg atoms, $M$ Rydberg atoms are positioned randomly inside a square area of side length $L$, which is considered a ``black-box''. Three additional atoms form the \textit{input site} and \textit{two output sites} and are placed outside the box, as shown in \fref{NN_models}~(a). Here, we consider only a single Rydberg $p$-excitation in the Rydberg aggregate, initially located on the \textit{input site}. The number of such excitations is conserved, allowing us to use a simplified notation for relevant many-body states $\ket{\pi_n}=\ket{ss...p..s}$, where only the $n$th atom is in the Rydberg $p$-state, and all others in the Rydberg $s$-state. 

Rydberg atoms interact among each other via long-range dipole-dipole interactions, governed by the aggregate Hamiltonian ($\hbar=1$) \cite{leonhardt2016orthogonal,ravets2015measurement}
\begin{equation}
	\sub{\hat{H}}{agg}=\sum_{n,m\neq n}\frac{(1-3\cos^2\theta_{nm})}{2}\frac{C_3}{|\mathbf{R}_{nm}|^3}\ket{\pi_n}\bra{\pi_m},
	\label{NN_H_agg}
\end{equation}
where $\mathbf{R}_{nm} = |\mathbf{R}_n - \mathbf{R}_m|$, with $\mathbf{R}_n=[x_n,y_n,z_n]^T$ the position of the $n$th Rydberg atom. The angle $\theta_{nm}$ is between the quantization axis and the separation vector $\mathbf{R}_{nm}$ linking the Rydberg atoms. A magnetic field is assumed to have removed components of the $p$-state other than $m_l=0$ in order to reach \eref{NN_H_agg}  \cite{leonhardt2016orthogonal}, with $\hbar m_l$ being the angular momentum projection on the quantisation axis. Dipole-dipole interactions now allow the single $p$-excitation to migrate on the Rydberg array, so that the latter forms a quantum network.
 
The density matrix $\hat{\rho}=\sum_{n,m}^N\rho_{n,m}\ket{\pi_n}\bra{\pi_m}$ of the system then evolves according to the master equation
\begin{equation}
	\dot{\hat{\rho}}=-i[\sub{\hat{H}}{agg}+\hat{H'},\hat{\rho}]+{\cal L}_{\hat{L}}[\hat{\rho}],
	\label{NN_Lindblad}
\end{equation}
where ${\cal L}$ is the super-operator written as ${\cal L}_{\hat{L}}[\hat{O}]=\hat{L}\hat{O}\hat{L}^\dagger-1/2\{\hat{L}^\dagger\hat{L},\hat{O}\}$
and we use $\hbar=1$ above and in the following.

The operators $\hat{H'}$ and $\hat{L}$ can arise from interactions between the Rydberg aggregate atoms and the atoms in the ultra-cold background gas, encapsulating the resultant disorder and decoherence:
\begin{eqnarray}
	\hat{H'}=\sum_n^N h'_n\ket{\pi_n}\bra{\pi_n},\label{NN_Hprime}\\
	\hat{L}=\sum_n^N l_n\ket{\pi_n}\bra{\pi_n}.
	\label{NN_L_d}
\end{eqnarray}
We will employ three variants for \eref{NN_Hprime}-\bref{NN_L_d} in this article: (i) Initially we set $h'_n=l_n=0$, such that only coherent dipole-dipole interactions \bref{NN_H_agg} are present. 
(ii) In a final section, we allow non-trivially varying but realistic $h'_n\neq 0$, $l_n\neq 0$ based on randomised background gas atom locations, as discussed in \aref{app:controllable_decoherence}.
(iii) As an interim model to systematically explore sensitivity of network tomography to decoherence, we retain only the decoherence rates $l_n$ from (ii) with $h'_n=0$, 
but rescale the $l_n$ to reach a target mean decoherence rates $\gamma$. 

We initialize the system with the single Rydberg $p$-excitation on the \textit{input site}, the red-ball in \fref{NN_models}~(a), the position of which is fixed at a distance $L$ to the left of the centre of the box. The initial aggregate state is thus
\begin{equation}
	\hat{\rho}(t=0)= \ket{\pi_1}\bra{\pi_1}.
\end{equation}
The excitation then can migrate to the \textit{output atoms} (violet-balls) via the \textit{black-box} region, due to the dipole-dipole interaction $\sub{\hat{H}}{agg}$, as shown in \fref{NN_confusion_mat}~(a). To describe this, the Lindblad master equation \bref{NN_Lindblad} is solved numerically, using the high-level programming language XMDS \cite{xmds:docu,xmds:paper}. Finally, we sample the $p$-excitation probability on the $k$th output atom at different coordinates $(\mathbf{r}_k^a)$, see \aref{app:data_collect}, with the measurement operator $\hat{F}_k = [\ket{p}\bra{p}]_k$, acting on output atom $k$, providing 
\begin{align}
	P_k(\mathbf{r}_k^a,\mathbf{X}_m,\hat{H'},\hat{L};t_{end})&=\text{Tr}_k[\hat{F}_k \hat{\rho}(t_{end})].
	\label{NN_input_datasets}
\end{align}
Here and in the following, we label the positions of the atoms inside the box by $\mathbf{X}_m$ and those of output atoms by $\mathbf{r}_k$. Probabilities $P_k(\mathbf{r}_k^a,\mathbf{X}_m,\hat{H'},\hat{L};t_{end})$ are recorded at 
a time $t_{end}$, as shown in \frefp{NN_confusion_mat}{b}, and form the input of the machine learning algorithm. 
Since training of a neural network requires large datasets, we take the two \textit{output atoms} to be systematically moved in small steps, providing discrete positions $\mathbf{r}_k^a$. The first output atom is moved along the $x-y$ plane in a semi-circular path of radius $L$, while the second output atom is moved simultaneously along the $y$-axis in a straight line at $z=L/2$, as sketched in \fref{NN_models}, allowing $200$ configuration in total for positions of the two output atoms.
The combined data of excitation probabilities on both output atoms, then provides a data-set of 400 probabilities for the machine learning models, see \aref{app:data_collect} for more details.

The numerical simulation is then repeated $10^4$ times, where the number of atoms ($M$) and their positions $\mathbf{X}_m^l$ are varied in each dataset $l$, with  $1\leq M \leq 4$ and positions 
$\mathbf{X}_m^l$ uniformly distributed in a square of edge length $L$. We keep $\hat{H'}=0$ and $\hat{L}=0$ for the moment. The generated datasets are then used for training of the machine learning algorithm and a separate dataset with $10^3$ elements is generated for testing, using $250$ cases for each $M$.  In the following sections, we will see how the measurement data, obtained from the \textit{output atoms} one and two, can provide us with useful information related to the quantum network as well as its decoherence by the environment using machine learning.
%
%%%%%%%%%%%%%%%%%%%%%%%%%%%%%%%%%%%
%
%%%%%%%%%%%%%%%%%%%%%%%%%%%%%%%%%%%
\section{Machine learning based multi-branch pipeline}
\label{NN_classification_regression}
%%%%%%%%%%%%%%%%%%%%%%%%%%%%%%%%%%%

In this section, we establish the multi-branch pipeline for
quantum network tomography of Rydberg arrays by machine learning, sequentially discussing
each stage of the pipeline: (i) \emph{classification},  where the system is classified based on the number of Rydberg atoms $M$ present in the black-box region using classification algorithms, (ii) \emph{multi-target regression}, where the algorithm infers the atomic coordinates, system Hamiltonian and Lindblad operators using artificial neural networks.

%%%%%%%%%%%%%%%%%%%%%%%%%%%%%%%%%%%
\subsection{Classification: number of atoms}
\label{sec:NN_class_results}
%%%%%%%%%%%%%%%%%%%%%%%%%%%%%%%%%%%
\begin{figure}[bth]
	\centering
	\includegraphics[width=0.99\linewidth]{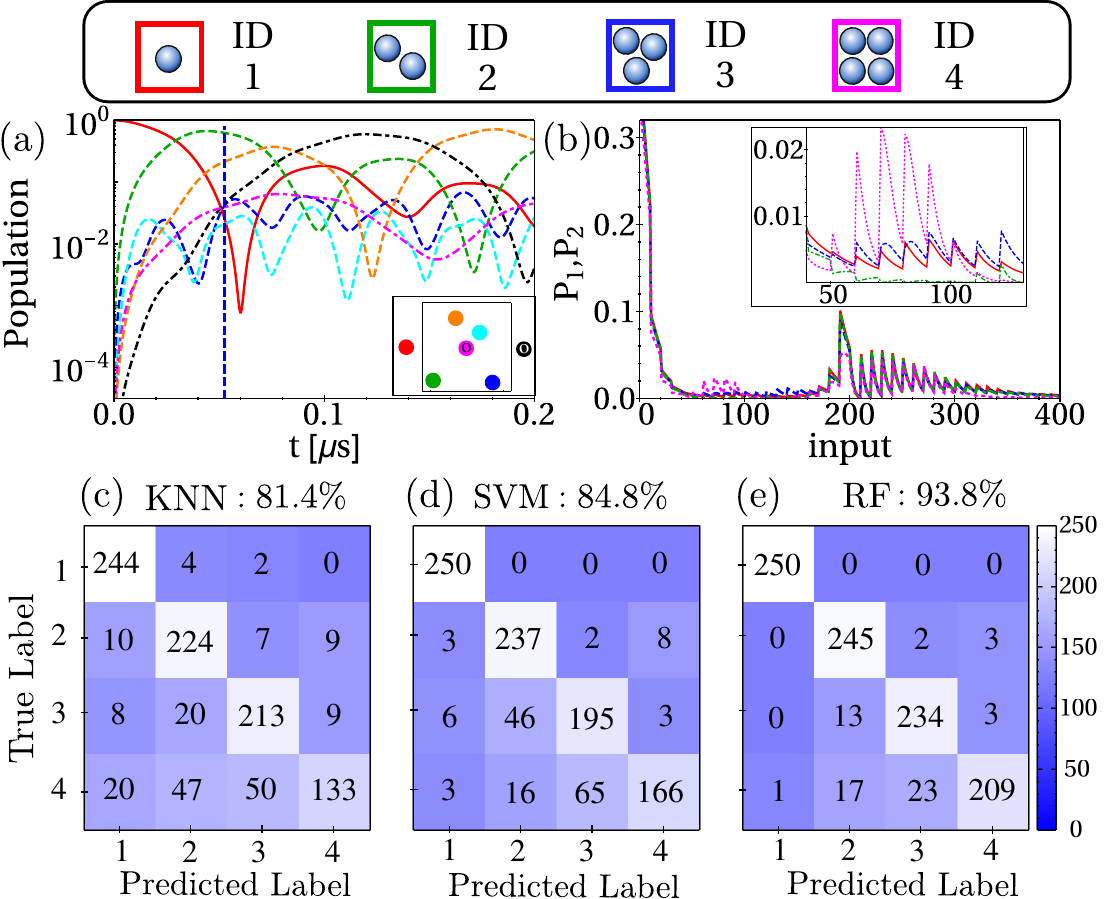}
	\caption{Determination of the number of network nodes. (a) Excitation transport in a Rydberg aggregate of $N=7$ atoms ($M=4$ in the box), showing population on the input site (red-solid), output atoms (black dot-dashed, magenta dot-dashed) and atoms inside the black-box (dashed). Positions of the atoms are shown in the inset with colour matching the corresponding lines. The blue dashed vertical line indicates $t_{end} = 0.05\mu$s for which output probabilities are extracted for later use. (b) Resultant input datasets at $t_{end} = 0.05\mu$s, described by \eref{NN_input_datasets} for $M=1$ (red solid), 2 (green dot-dashed), 3 (blue dashed), and 4 (magenta dotted) for a single realisation of random atom positions in the box. The inset shows a zoom on the low probability structure. (c-e) Confusion matrix for inference of the number of nodes $M$, resulting from different classification algorithms (c) K-Nearest Neighbor (KNN), (d) Support Vector Machine (SVM) and (e) Random Forest (RF), here the label corresponds to $M$, visualized in the top-most panel.
The box size of $L=10\mu$m is the same for (a)-(e).  }
	\label{NN_confusion_mat}
\end{figure}
In the first stage of the pipeline, we classify the system according to the number of Rydberg atoms inside the \textit{black-box} region. We provide the datasets $P_k(\mathbf{r}_k^a,\mathbf{X}_m^l,\hat{H'}_l,\hat{L}_l;t_{end})$ containing excitation probability on the output atoms to a set of classification algorithms, which infer the number of Rydberg atoms within the box. We have considered four classes ($N_{ID}=4$) with each class (ID) characterised by the number of atoms $M$, as shown at the top of \fref{NN_confusion_mat}. The variation within the input dataset for each class can be seen in \fref{NN_confusion_mat}~(b) for a single realisation. The figure shows excitation probabilities on both output atoms, where inputs $1-200$ are excitation probabilities at the output site $1$ ($P_1$), while inputs $201-400$ correspond to output site $2$ ($P_2$). Coarsely, the input datasets for each class appear similar, but in detail the relative probability differences between the input datasets of each class can be large, as shown in the inset of \fref{NN_confusion_mat}~(b). These minor distinctions can play an important role in the identification of each class.
\begin{figure}[htb]
	\centering
	\includegraphics[width=0.99\linewidth]{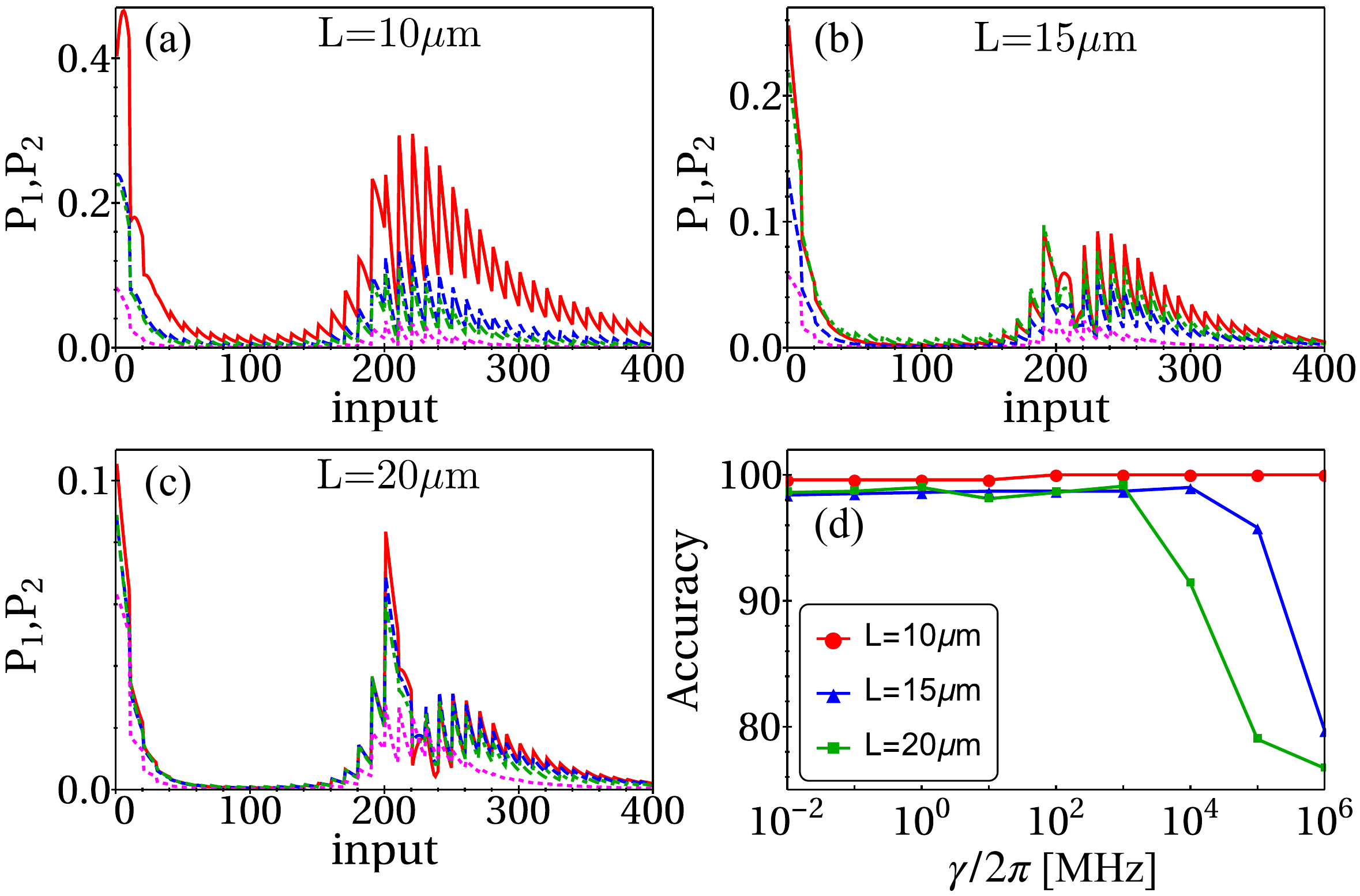}
	\caption{Input data sets for different black box sizes, 
	with atoms subject to finite mean decoherence $\gamma/2\pi=10^2$ MHz.
Input datasets are shown for (a) $L=10\mu$m, (b) $L=15\mu$m and (c) $L=20\mu$m similar to \fref{NN_confusion_mat}, but here with $\gamma>0$. (d) Accuracy of atom number classification by Random Forest with an increase in mean decoherence $\gamma$ for $L=10\mu$m (red circle), $L=15\mu$m (blue triangle) and $L=20\mu$m (green square). See \aref{app:decoh_effect_pop} for corresponding excitation transport dynamics at $1\leq\gamma/2\pi \leq 10^3$ MHz ($\hbar=1$).}
	\label{NN_accuracy_drop_plot}
\end{figure}
Here, we focus on three classification algorithms: (i) Support vector machine (SVM), (ii) Random forest (RF), and (iii) K-Nearest neighbor (KNN). A SVM establishes the optimal decision boundary that divides the classes and optimises the distance between the optimal decision boundary and the closest data point from each class \cite{hearst1998support}. The decision boundary for data with p features or classes is a
$(p - 1)$-dimensional hyperplane.

In contrast to the SVM, a random forest (RF) is an ensemble approach that combines
the predictions of several predictors to enhance its performance \cite{belgiu2016random}. The final prediction is determined by the mean of all individual predictions. Typically, an ensemble approach outperforms its constituent predictors, particularly when the constituent predictors make diverse errors. Random forests consist of decision trees that have been trained on random subsets of samples and features to increase their performance.

K-Nearest neighbor (KNN) is one of the simplest machine learning algorithms based on the similarity between a new case and pre-existing cases. It places each new case into a category that is most similar to the pre-existing categories \cite{mucherino2009k}. The KNN algorithm stores all available data and classifies a new data point based on its similarity to the existing data. This implies that as new data becomes available, the KNN algorithm can readily classify it into an appropriate category. KNN is a non-parametric method, hence it does not make assumptions about the underlying data.

In \fref{NN_confusion_mat}, the results obtained from three classifiers are shown, namely, (i) K-Nearest Neighbor (KNN), (ii) Support Vector Machine (SVM) and (iii) Random Forest (RF) classifier. These classifiers are trained with $N_{\mathrm{train}}=10^4$ datasets and tested on $N_{\mathrm{test}}=10^3$ cases, all with equal contributions from each $M\in\{1,2,3,4\}$. We use a $N_{ID}\times N_{ID}$ confusion matrix for a graphical representation of the performance of the classifier, shown in \fref{NN_confusion_mat}~(c-e). A confusion matrix shows a comparison between predicted IDs and actual IDs, with the diagonal elements representing correct predictions ($N_\mathrm{correct}$=No. of instances with predicted ID=actual ID) and the off-diagonal elements indicating the wrong ones ($N_\mathrm{incorrect}$=No. of instances with predicted ID$\neq$actual ID). It can be seen in \fref{NN_confusion_mat}~(e), that the Random Forest classifier presents the best results with an accuracy ($=N_\mathrm{correct}/N_{\mathrm{test}}$) of $93.8\%$, while KNN and SVM produce an accuracy of $81.4\%$ and $84.8\%$ as shown in \fref{NN_confusion_mat}~(c) and (d), respectively. 
 
So far, we have considered coherent dynamics in the system by setting the decoherence $\hat{L}=0$ in \eref{NN_Lindblad}, using $l_n=0$ in \eref{NN_L_d}. Now, in order to analyse the robustness of classification towards environmental noise, we systematically vary the decay parameters $l_n$ in the system to induce controlled decoherence of mean strength $\gamma$, given by
\begin{equation}
	\gamma/2\pi={\cal N} \overline{ \gamma_{nm}}/2\pi.
	\label{NN_gamma}
\end{equation} 	
Here, $\gamma_{nm}/2\pi=|l_n|^2+|l_m|^2-2\text{Re}[l_nl_m^*]$ are all decoherence rates in the black box, acting as ${\cal L}_{\hat{L}}[\hat{\rho}]=\sum_{n,m} \gamma_{nm} \rho_{n,m}$ and discussed in more detail in \aref{app:controllable_decoherence}. In the expression above, $\overline{\cdots}$ denotes the average over all pairs of sites in a single realisation. We then use the scaling parameter ${\cal N}$ to conveniently tune the mean decoherence strength relative to other energy scales. In \sref{sec:NN_Heff_Leff} we shall employ a fully realistic model instead.
 
  By comparing \frefp{NN_accuracy_drop_plot}{a} with \frefp{NN_confusion_mat}{b}, we can see how the input dataset is modified by decoherence at rate $\gamma/2\pi=10^2$ MHz for $L=10 \mu$m. Each class now results in a more significantly distinct input dataset. Therefore, near $\gamma/2\pi=10^2$ MHz, the random forest classifier in \frefp{NN_accuracy_drop_plot}{d}, even shows a slight increase in accuracy of up to $99\%$ compared to $93.8\%$ at $\gamma=0$, for a box-length $L=10\mu m$. This however breaks down at extremely high decoherence $\gamma/2\pi=10^4$ MHz due to identical inputs for $L=10\mu$m and $20\mu$m, see \aref{app:decoh_effect_pop} (\fref{pop_decoh}). A similar classification problem has been investigated in \rref{chong2022machine} for atoms arranged in pre-fixed configurations, but here we extend this to a more complex scenario, by randomising the distribution of Rydberg atoms across the $x-y$ plane. In addition, we incorporate long-range dipole-dipole interactions with decoherence and limit the datasets to the information from two output atoms at a given time instead of the complete time-evolution of the wavefunction.

%%%%%%%%%%%%%%%%%%%%%%%%%%%%%%%%%%%
\subsection{Regression: quantum network reconstruction}
\label{sec_NN:regression_QNR}
%%%%%%%%%%%%%%%%%%%%%%%%%%%%%%%%%%%
\begin{figure}[htb]
	\centering
	\includegraphics[width=0.99\linewidth]{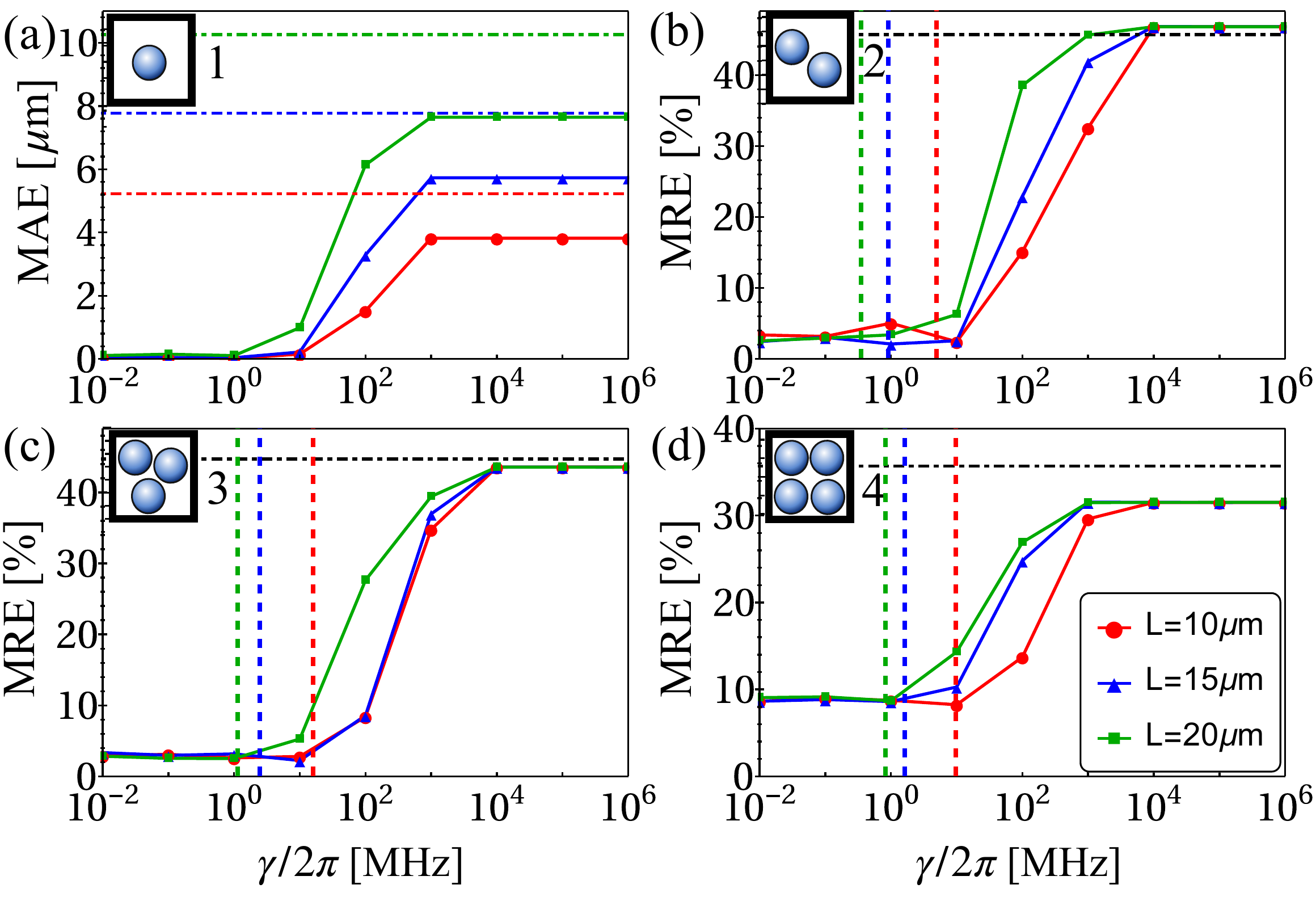}
	\caption{ Finding Rydberg locations with the neural network. (a) Mean absolute error (MAE) between predicted position and actual position for $M=1$ and box sizes $L=10\mu m$ (red circle), $L=15\mu$m (blue triangle) and $L=20\mu$m (green square). Horizontal dot-dashed lines indicate the maximum MAE for each box size in the same colors, averaged over $10^3$ random realisations, see text. (b)-(d) Mean relative error (MRE) described by \eref{NN_mae_set} for $M=2,3,4$, respectively as sketched in the insets, for box sizes as in (a). Horizontal black dot-dashed lines indicate the maximum MRE, see \eref{NN_mae_max}. Vertical dashed lines indicate average dipolar interaction strengths from $\sub{\hat{H}}{agg}$, with colors allocated to box sizes as in other panels. A shared legend is provided in (d).}
	\label{NN_mae_pos_plot}
\end{figure}
The second stage of the pipeline utilises a multi-target regression algorithm to locate the Rydberg atoms in the \textit{black-box} reg,ion. We use the input datasets obtained from the two output atoms $P_{1,2}(\mathbf{r}_{1,2 }^a,\mathbf{X}_m^l,\hat{H'}_l,\hat{L}_l;t_{end})$, as shown in \fref{NN_confusion_mat}~(b) and \fref{NN_accuracy_drop_plot}~(a-c), along with the information  about the number of atoms $M$ gained by the previous stage to separately tackle each $M$. The artificial neural network is then trained with $10^4$ datasets, also including the known positions of the Rydberg atoms within the black box. This enables it to later predict the coordinates of the Rydberg atom at the output of the neural network for new unseen test data. Since each $M$ requires the prediction of a different number of coordinates, we employ neural networks with a different number of output neurons for each $M$.

To assess the performance of the neural network, we consider the mean relative error ($MRE$) defined as
\begin{equation}
	MRE_p = \frac{1}{M} \sum_{n=2}^{M+1}\frac{(|\mathbf{R}_n^{p,pred}-\mathbf{R}_n^p|)}{d_{mean}},
	\label{NN_mae_set}
\end{equation}
where $M=N-3$ is the number of Rydberg atoms in the box and $d_{mean}$ is the mean distance between the Rydberg atoms inside the box. $\mathbf{R}_n^p$ is the actual position of the $n$th Rydberg atom in the $p$th dataset, while $\mathbf{R}_n^{p,pred}$ is the position predicted by the neural network.
Since predicted positions are not labelled, we allocate the one closest to the $n$th Rydberg atom. For the single atom case, $M=1$, we switch the assessment parameter from the mean relative error to the mean absolute error ($MAE$) defined as $MAE_p = \sum_{n=2}^{M+1}|\mathbf{R}_n^{p,pred}-\mathbf{R}_n^p|$. 

Altogether, we train the multi-target artificial neural network with $10^4$ datasets for each $M$, including decoherence ranging from $\gamma/2\pi=10^{-2}$ MHz to $10^6$ MHz, relative to a mean strength of $\overline{\langle\sub{\hat{H}}{agg}\rangle}/2\pi\approx [3.3,15.9,5.5]\times 10^3/L^3$ MHz $\mu$m$^3$ for $M=\{2,3,4\}$. Here $\overline{\cdots}$ is the average over all site indices $n,m$ and all realisations. We show the variation of the errors $MRE$ (and $MAE$) with respect to decoherence strength $\gamma/2\pi$ in \fref{NN_mae_pos_plot}, for three different box lengths $L=10,15,20$ $\mu$m. For this we used $10^3$ test datasets in each case.
Training and testing is done separately for each surveyed decoherence strength $\gamma$.
The horizontal black dot-dashed lines in (b-d) indicate the worst expected $MRE$, corresponding to a completely random prediction. This limit is determined by averaging over $P_N=10^3$ random and uncorrelated realisations of both, ``fake predicted'' and ``fake actual'' atoms:
\begin{equation}
	\sub{MRE}{max} = \frac{1}{P_N M} \sum_{p,p'}^{P_N}\sum_{n,n'=2}^{M+1}\frac{(|\mathbf{R}_{n'}^{p'}-\mathbf{R}_n^p|)}{d_{mean}},
	\label{NN_mae_max}
\end{equation}
where $n$ and $n'$ are sorted based on proximity of the $n$th atom and $n'$th atom in the $p$th and $p'$th dataset, respectively, as discussed below \eref{NN_mae_set}. It is then clear that one should consider cases with 
 $MRE\approx \sub{MRE}{max}$ as complete failure of position reconstruction.

Compared to the classification algorithm in \fref{NN_accuracy_drop_plot}, the regression algorithm breaks down at a lower threshold decoherence rate $\gamma$. By threshold for breakdown, we refer to those values of $\gamma$ where the $MRE$ approaches $MRE_{\mathrm{max}}$ in \fref{NN_mae_pos_plot}. We also observe an earlier 
threshold where the $MRE(\gamma)$ starts to increase from its initial low value that is independent of $\gamma$. This appears to be set by the mean strength of the dipolar interactions in the system, $\overline{\langle \sub{\hat{H}}{agg} \rangle }$, shown with vertical dashed lines in \fref{NN_mae_pos_plot} for different box-lengths $L$. Reconstuction worsens as soon as $\gamma$ exceeds this scale, $\gamma\geq \overline{\langle \sub{\hat{H}}{agg} \rangle}$. We discuss the reasons for this in \aref{app:decoh_effect_pop} (\fref{pop_decoh}). Furthermore, the increase of the minimal error from $2\%$ in \fref{NN_mae_pos_plot}~(c) to $8\%$ in \fref{NN_mae_pos_plot}~(d) upon increasing the number of atoms inside the box from $M=3$ to $4$ indicates that our neural network architecture is challenged more strongly by larger systems. However we find that these errors can be again reduced by increasing the training set size.

%%%%%%%%%%%%%%%%%%%%%%%%%%%%%%%%%%%
\subsection{Regression: system and environment}
\label{sec:NN_Heff_Leff}
%%%%%%%%%%%%%%%%%%%%%%%%%%%%%%%%%%%

We now adapt the artificial neural network developed in the previous section to a new set of objectives: predicting the matrix elements of the system-environment interaction Hamiltonian $\hat{H'}$ and Lindblad operator $\hat{L}$ of the system, described by \eref{NN_Heff} and \eref{NN_Leff}, respectively in \aref{app:controllable_decoherence}. In the companion article \cite{mukherjee:modelbuilding}, we suggest a more general approach, for which we demonstrate a realistic experimental platform for demonstrations here, based on  Rydberg array. However, instead of predicting a completely arbitrary Hamiltonian and decay operators as in \cite{mukherjee:modelbuilding}, here, we work around a back-bone $\sub{\hat{H}}{agg}$, which depends on randomized positions of atoms and can be obtained in the previous step, while only corrections $\hat{H}'$ and $\hat{L}$ capture system-environment interactions.

For a realistic source of these corrections, the quantum network is considered to be immersed in a cloud of background atoms that can be optically manipulated to generate controllable decoherence for the transport of the  $p$-excitation, as discussed in \cite{David:Rydagg,schempp:spintransport,wuester:immcrad,genkin:markovswitch,Mukherjee_binding_dephasing_PhysRevA,mukherjee2020two,Mukherjee_excitons_polaritons}. 
In essence, quantum non-demolition (QND) measurements by the gas environment cause decoherence and energetic disorder.
Interactions causing this are artificially induced and thus controllable \cite{David:Rydagg}. The $h'_n$ and $l_n$ depend on the random positions of background gas atoms and parameters of optical fields, as  summarized in \aref{app:controllable_decoherence}. This results in a wide range of values for Hamiltonian matrix elements and dephasing rates, with which one can test and validate our neural network model. 

\begin{figure}[htb]
	\centering
	\includegraphics[width=0.99\linewidth]{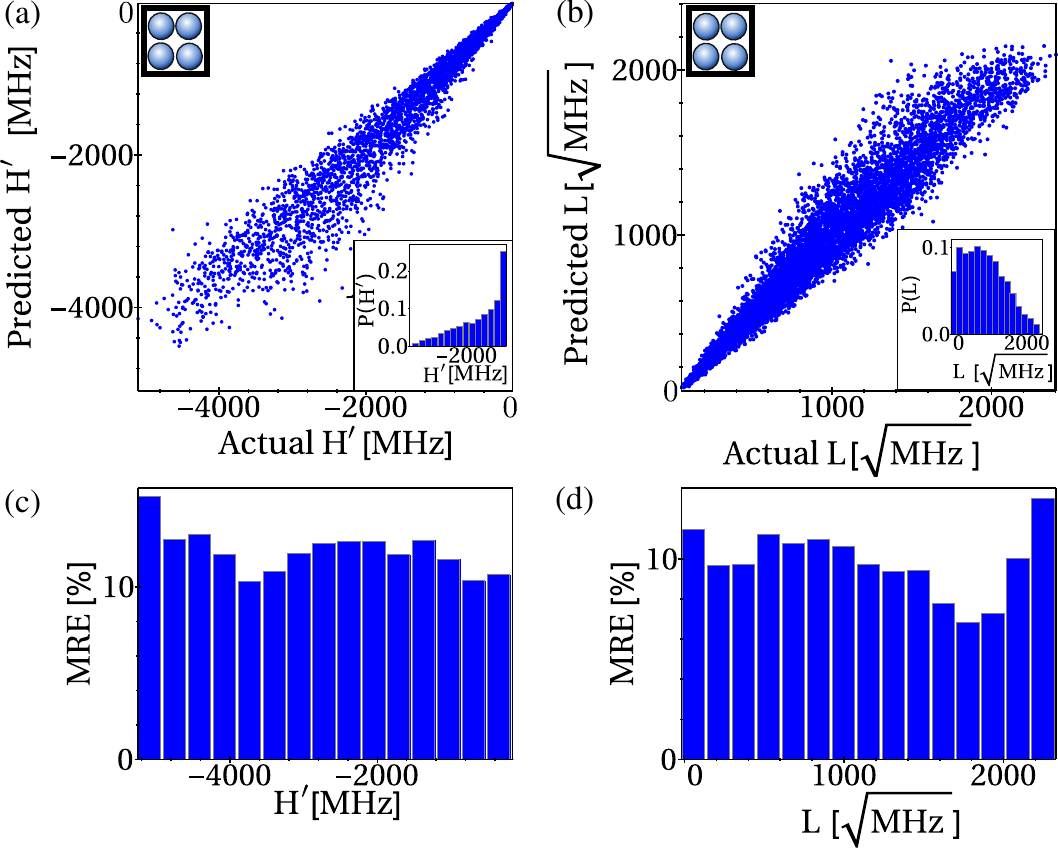}
	\caption{Machine learning of system and environment properties. We show a
comparison between predicted and actual matrix elements of (a) $\hat{H}'$ and (b) $\hat{L}$ defined in \eref{NN_Heff} and \eref{NN_Leff}, respectively. A histogram of the distribution of the respective effective operators is shown in the inset. EIT parameters, see \aref{app:controllable_decoherence}, are: $1\leq \Omega_p/2\pi\leq 13$ MHz, $\Omega_c/2\pi=30$ MHz, $\Gamma_p/2\pi=6.1$ MHz, $N_{bg}=6,400$, $\rho_{bg}=1.6\times 10^{13}$ m$^{-2}$ and $\ket{u}\equiv\ket{38s}$. Distribution of mean relative error (MRE) for (c) $\hat{H}'$ and  (d) $\hat{L}$.}
	\label{NN_David_eff}
\end{figure}
The matrix elements and dephasing rates are randomised through two experimentally accessible handles, the random Rydberg atom locations within the box and a probe light Rabi frequency $\Omega_p$ (see \aref{app:controllable_decoherence}) that is uniformly randomized in the range $1\leq \Omega_p/2\pi\leq 13$ MHz.
The results obtained are shown in \fref{NN_David_eff}~(a) and (b), where the blue dots have coordinates $(x,y)\equiv(h_{n}^{\prime\mathrm{predicted}},h_{n}^{\prime\mathrm{actual}})$ and $(x,y)\equiv(l_{n}^{\mathrm{predicted}},l_{n}^{\mathrm{actual}})$, respectively, such that dots closer to the diagonal indicate better results ($h_{n}^{\prime\mathrm{predicted}}=h_{n}^{\prime\mathrm{actual}})$. In \fref{NN_David_eff}~(c-d), we show the distribution of mean relative error ($\mathrm{MRE}(H_{nm})=(|h_{n}^{\prime\mathrm{predicted}}-h_{n}^{\prime\mathrm{actual}}|)/h_{n}^{\prime \mathrm{actual}}$) across the range of energies in $\hat{H'}$ and similarly for $\hat{L}$, which demonstrates that the neural network can predict the matrix elements of the operators with a mean relative error of $\approx 10\%$, consistent throughout the energy range. In \cite{mukherjee:modelbuilding}, the performance of the neural network has been tested on completely unconstrained Hamiltonians and decay operators, with results suggesting that the neural network can be generalized for a complete reconstruction of arbitrary quantum network.

%%%%%%%%%%%%%%%%%%%%%%%%%%%%%%%%%%%
\section{Conclusion and outlook}
\label{concl}
%%%%%%%%%%%%%%%%%%%%%%%%%%%%%%%%%%%

We have demonstrated quantum network tomography of simple quantum systems using a machine learning algorithm. We designed a multi-branch pipeline, aiming to provide key information about a quantum network and the strength and character of disorder and decoherence in the system. We apply this technique specifically to an embedded Rydberg aggregate model, focussing on realistic parameters that should be accessible in state-of-the-art experiments. The first stage of our algorithm successfully classified the number of atoms in the Rydberg aggregate and thus the number of nodes in the network. 
The second stage contained a separate branch, each handling a different number of nodes.
Here, an artificial neural network predicts the position of the atoms and details of disorder and decoherence. We have demonstrated that it can successfully do so, based on measurements of the probability for an excitation to be transported through the network and reaching a single target output site. This information is only required for a single snapshot in time.

Future work should validate the utility of our machine learning algorithm by combining theory with state-of-the-art experiments, determining the essential properties and characteristics of the system based on experimentally measured data, using a theory trained neural network. A similar approach could also be leveraged towards quantum state tomography of Rydberg arrays, where we reconstruct a complex entangled quantum state initialised initially through the resultant transport characteristics when coupled to a larger system, based on a similar
neural network architecture. Ultimately, the applications of our multi-branch pipeline may extend to quantum network tomography on complex molecular systems like photosynthetic molecular aggregates or quantum devices like SQUID based quantum computing architectures, where the approach may provide key information about complex structures and system-environment interactions.

%%%%%%%%%%%%%%%%%%%%%%%%%%%%%%%%%%%
\acknowledgments
We gladly acknowledge interesting discussions with Abhijit Pendse, Ritesh Pant and thank the Max-Planck society for financial support under the MPG-IISER partner group program as well as the Indo-French Centre for the Promotion of Advanced Research - CEFIPRA. K.M.~acknowledges the Ministry of Education for the Prime Minister's Research Fellowship (PMRF). We acknowledge support by the European Union's Horizon Europe research and innovation program under the project MLQ under Marie Skłodowska-Curie grant agreement number 101120240. Work was also supported by the CNRS through the EMERGENCE@INC2024 project DINOPARC and by the French National Research Agency under the Investments of the Future Program project ANR-21-ESRE-0032 (aQCess) and the Institut Universitare de France
(IUF).
%%%%%%%%%%%%%%%%%%%%%%%%%%%%%%%%%%%

%%%%%%%%%%%%%%%%%%%%%%%%%%%%%%%%%%%
%%%%%%%%%%%%%%%%%%%%%%%%%%%%%%%%%%%
\appendix
\beginsupplement
%%%%%%%%%%%%%%%%%%%%%%%%%%%%%%%%%%%
%%%%%%%%%%%%%%%%%%%%%%%%%%%%%%%%%%%	
\section{Controllable decoherence and disorder in Rydberg array}
\label{app:controllable_decoherence}
%%%%%%%%%%%%%%%%%%%%%%%%%%%%%%%%%%%	
To explore a setting where artificial controllable decoherence can be set in experiments to test the neural network reconstruction of it, we assume that the EIT is selectively implemented on the background gas in the black-box region \cite{guenter:EITexpt,David:Rydagg}. 

One can encapsulate all the system-environment interactions involving van der Waals interactions and EIT into a set of effective operators in the Rydberg state-space, which can take the form \cite{David:Rydagg} 
\begin{eqnarray}
	h'_n &=& \sum_\alpha\frac{\Omega_p^2}{\Omega_c^2} \frac{\bar{V}_{n\alpha}(\mathbf{X}_n,\mathbf{x}_\alpha)}{1+(\bar{V}_{n\alpha} (\mathbf{X}_n,\mathbf{x}_\alpha)/V_c)^2},\label{NN_Heff}\\
	l_n^{(\alpha)} &=&  \frac{\Omega_p}{\sqrt{\Gamma_p}}\frac{1}{i+V_c/\bar{V}_{n\alpha}(\mathbf{X}_n,\mathbf{x}_\alpha)},\label{NN_Leff}
\end{eqnarray}
where $\bar{V}_{n\alpha}=V_{n\alpha}^{(rp)}(\mathbf{X}_n,\mathbf{x}_\alpha)+\sum_{m\neq n}V_{m\alpha}^{(rs)}(\mathbf{X}_m,\mathbf{x}_\alpha)$ accounts for the van der Waals interactions $V_{n\alpha}^{(ra)}(\mathbf{X}_n,\mathbf{x}_\alpha)=C_{\eta(a)}/|\mathbf{X}_n-\mathbf{x}_\alpha|^{\eta(a)}$ with $\eta(a)=6,4$ for $a=s,p$, respectively. Here $\mathbf{X}_n$ and $\mathbf{x}_\alpha$ indicate the position of the $n$th Rydberg atom and the $\alpha$th background gas atom, respectively. $\Omega_{p}$ and $\Omega_c$ are probe and coupling Rabi frequencies of EIT, $V_c=\Omega_c^2/(2\Gamma_p)$. For simplicity, we predict an effective decay rate from the algorithm which can be approximately written as $l_n \approx \sum_\alpha l_n^{(\alpha)}$.

The master equation obtained from \eref{NN_Lindblad} by inserting $\hat{H'}$ and $\hat{L}$ can be written in the form
\begin{align}
	\dot{\rho}_{nm} =& \sum_k i(W_{km} \rho_{nk} - W_{nk} \rho_{km}) \nonumber\\&+ i(h'_m-h'_n+\epsilon_{nm})\rho_{nm}-\gamma_{nm}\rho_{nm}/2,\label{general_N_eom}
\end{align}
where $W_{nm}=(1-3\cos^2\theta_{nm})C_3/2|\mathbf{R}_{nm}|^3$ are dipole-dipole interactions defined in \eref{NN_H_agg} 
and, $\gamma_{nm}=|l_n|^2+|l_m|^2-2\text{Re}[l_nl_m^{*}]$ can be derived as follows
\begin{align}
	{\cal L}_{\hat{L}}[\hat{\rho}] &= \hat{L}\hat{\rho}\hat{L}^\dagger-1/2\{\hat{L}^\dagger\hat{L},\hat{\rho}\}, \nonumber\\
	&=l_nl_m^* \rho_{nm} - 1/2 (|l_n|^2+|l_m|^2)\rho_{nm}, \nonumber\\
	&=(i\text{Im}[l_nl_m^*] - 1/2(|l_n|^2+|l_m|^2-2\text{Re}[l_nl_m^{*}]))\rho_{nm},\nonumber\\
	&=(i\epsilon_{nm}-1/2\gamma_{nm})\rho_{nm}.
\end{align}

In the generation of each data set in \sref{sec:NN_Heff_Leff}, we randomise the EIT parameter $\Omega_p$ in the range $1\leq \Omega_p \leq 13$ MHz along with the Rydberg atomic positions, resulting in a wide range of values in the operators for testing and validating our neural network model. Note that prior to  \sref{sec:NN_Heff_Leff}, we use artificially engineered mean decoherence rates with scaling parameter, which are not directly governed by \eref{NN_Heff} and \bref{NN_Leff}.
%%%%%%%%%%%%%%%%%%%%%%%%%%%%%%%%%%%
\section{Design of cold atom benchmark platform and data collection}
\label{app:data_collect}
%%%%%%%%%%%%%%%%%%%%%%%%%%%%%%%%%%%

The following conditions constrain how we distribute Rydberg atoms inside the \textit{black-box region}: (i) the minimum distance between any two atoms is kept at $3.1 \mu$m (e.g.~the blockade radius for $\nu=43$ and excitation laser linewidth $3$ MHz, and (ii) the atoms are distributed evenly in the box, for example if $M=4$, the box is divided into four quadrants with each atom distributed randomly in one of the quadrants. 

We place the first output atom ($P_1(\mathbf{r}_1^a,\mathbf{X}_m^l,\hat{H'}_l,\hat{L}_l;t_{end})$) at 20 equidistant locations along the rim depicted in \frefp{NN_models}{a} and 10 successive positions radially outwards with $dr=0.2 \mu$m for each site on the rim to generate a dataset of 200 data points, as a part of the input to the neural network. In order to position the second output atom $P_2(\mathbf{r}_2^a,\mathbf{X}_m^l,\hat{H'}_l,\hat{L}_l;t_{end})$, we choose 20 evenly spaced locations at $z=r$ with $x$ ranging from $-r $ to $r $ with $y=0$, as well as 10 successive positions on each of those points along the $z-$axis with $dz=0.2\mu$m. The two output atoms are moved simultaneously, and generates a combined input data set of 400 data points from the two output atoms, where the first 200 points corresponds to $P_1$ and the next 200 points relates to $P_2$, which are shown in \fref{NN_confusion_mat}~(b). Finally, using a set of machine learning classification algorithms, we classify the system according to the number of Rydberg atoms in the network using the datasets $\{P_1(\mathbf{r}_1^a,\mathbf{X}_m^l,\hat{H'}_l,\hat{L}_l;t_{end}),P_2(\mathbf{r}_2^a,\mathbf{X}_m^l,\hat{H'}_l,\hat{L}_l;t_{end})\}$. For each example of $L=10 \mu$m, $15 \mu$m, and $20 \mu$m, we scan through different $t_{end}$ and found an optimal time sample for the training of the neural network at $t_{end}=0.05 \mu$s, $0.08 \mu$s, and $0.1\mu$s, respectively.

\begin{figure}[htb]
	\centering
	\includegraphics[width=0.99\columnwidth]{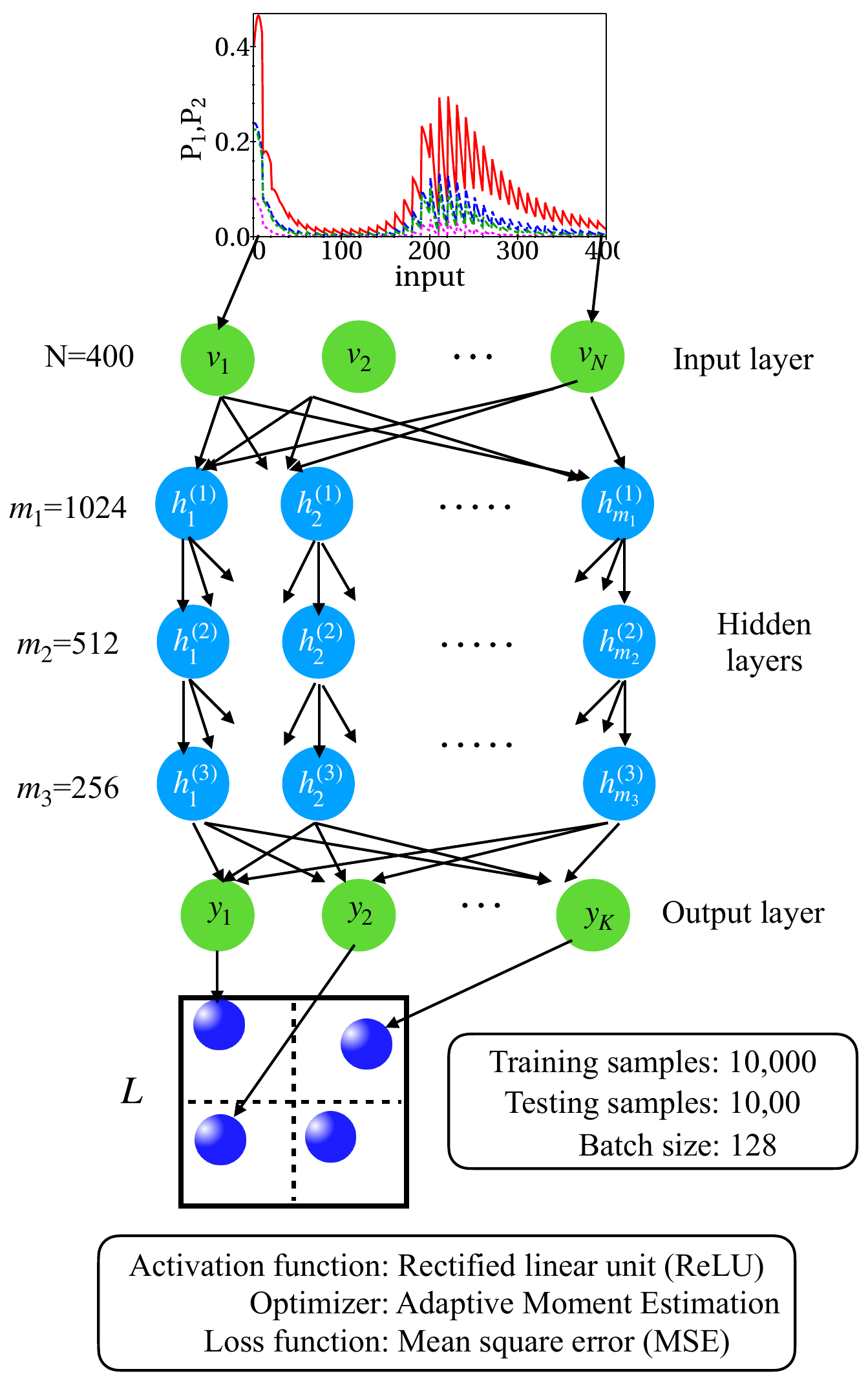}
	\caption{Schematic representation of artificial neural network (ANN) with an
		input layer consisting of $N=400$ input neurons ($\mathbf{v}$), fed with excitation probabilities of output sites similar to \frefp{NN_accuracy_drop_plot}{a} and shown at the top. This is followed by
$\sub{n}{hl}=3$  hidden layers, with layer $(i)$ consisting of $\sub{m}{hl}^{(i)}$ hidden
		neurons ($\mathbf{h}^{(i)}$) and an output layer with $K$ output neurons ($\mathbf{y}$), which provide either atomic positions or Hamiltonian or Lindblad operator matrix element. Information is processed
		from top to bottom, as indicated by the arrows. }
	\label{NN_structure}
\end{figure}
%
%%%%%%%%%%%%%%%%%%%%%%%%%%%%%%%%%%%
\section{Design of Artificial Neural Network}
\label{app:NN_design}
%%%%%%%%%%%%%%%%%%%%%%%%%%%%%%%%%%%

Once the network is classified based on the number of atoms as described in \sref{sec:NN_class_results}, in the next stage, we use artificial neural network with multi-target regression to predict the location of those atoms. In general, a artificial neural network consists of three sections, for which we are using the following specifications:
\begin{enumerate}
	\item Input layer: Our input layer consists of $400$ neurons, where the measurement data of Rydberg $p$-excitation on the output atoms, $P_k^l(\mathbf{r}_k^a,\mathbf{X}_m^l,\hat{H'}_l,\hat{L}_l;t_{end})$ for $k=1,2$ are provided, obtained as discussed in \aref{app:data_collect}.
	\item Hidden layer: We have typically included $\sub{n}{hl}=3$ hidden layers, where the $i$th layer in the forward direction has $\sub{m}{hl}^{(i)}=2^{11-i}$ neurons. All the neurons are connected to each other with dropout = 0, and activated using Rectified linear unit (ReLU) activation function. We have used mean squared error (MSE) as our loss function and Adaptive Moment estimation (ADAM) as optimizer.
	\item Output layer: We vary the number of neurons at the output layer based on data type required for the prediction. For instance, to predict the location of four atoms inside the 2D box, there are $8$ coordinates $(x_1,y_1),\cdots,(x_4,y_4)$ constituting $8$ real numbers, requiring $8$ neurons in the output layer.	
\end{enumerate}
The neural network configured like this was trained with $10^4$ training datasets in all cases for typically $10^3$ epochs (iterations for the loss function optimisation during training).
To prevent underfitting or overfitting, the number of epochs was adjusted from that value case by case. 
%%%%%%%%%%%%%%%%%%%%%%%%%%%%%%%%%%%
\section{Effect of decoherence on populations}
\label{app:decoh_effect_pop}
%%%%%%%%%%%%%%%%%%%%%%%%%%%%%%%%%%%

In {\sref{NN_classification_regression}, we have seen that there is a critical level of decoherence beyond which the neural network no longer performs. Here, we examine the cause of this. In \fref{pop_decoh}~(a-d), we can see that the coherent excitation transport approaches a steady state with an increase in decoherence. Furthermore, the amount of excitation passing through the boxed atoms decreases, see the green dashed line in \fref{pop_decoh}~(a-d), making it impossible for us to gather any relevant information regarding the location of atoms. As a result, each simulation generates a similar dataset, and the neural network can no longer accurately locate the atoms. However this is not the case for classification, as can be seen from the slight increase in performance of  the Random forest classifier in \fref{NN_accuracy_drop_plot}~(d), until a certain regime because the variation in atom number results in a different steady state for each case.
	Additionally, we can see in \fref{pop_decoh}~(e-f), at extremely high decoherence, for example, at $\gamma/2\pi = 10^6$ MHz, how different numbers of atoms $M$ can result in identical input datasets, such as $M=2$ and $3$ in \fref{pop_decoh}~(e) and $M=1$-$3$ in \fref{pop_decoh}~(f), which causes the drop in the accuracy due to degenerate input datasets for classification algorithms.

\begin{figure}[htb]
	\includegraphics[width=\linewidth]{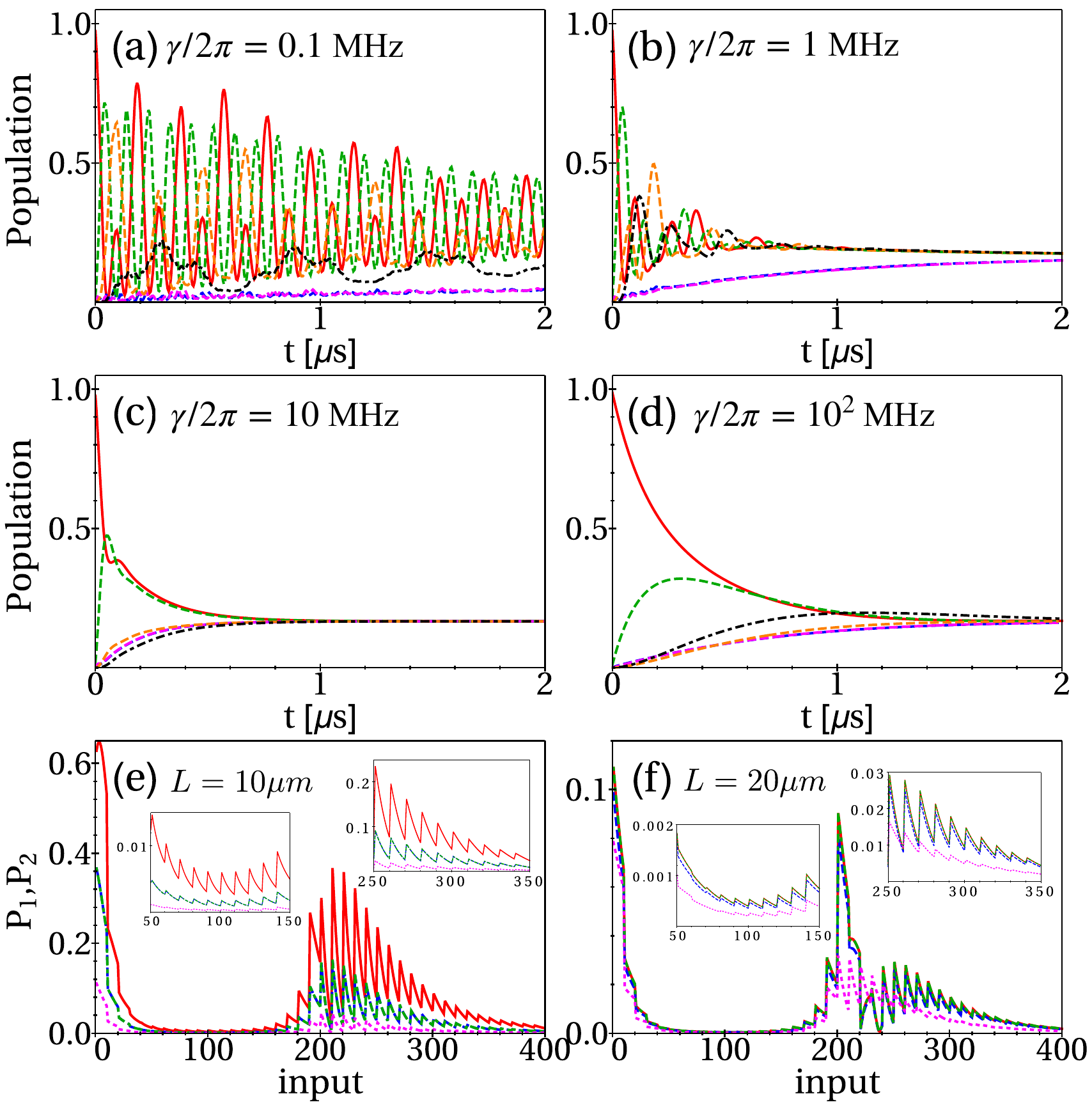}
	\caption{Excitation transport similar to \fref{NN_confusion_mat}~(a) with $L=10\mu$m but for (a) $\gamma/2\pi = 0.1$ MHz, (b) $\gamma/2\pi = 1$ MHz, (c) $\gamma/2\pi = 10$ MHz and (d) $\gamma/2\pi = 1,00$ MHz. Below we show input datasets similar to \fref{NN_accuracy_drop_plot}~(a) and (c) corresponding to box size (e) $L=10\mu$m and (f) $L=20\mu$m for $\gamma/2\pi=10^6$ MHz with  $M=1$ (red solid), 2 (green dot-dashed), 3 (blue dashed), and 4 (magenta dotted)}
	\label{pop_decoh} 
\end{figure}
%

%%%%%%%%%%%%%%%%%%%%%%%%%%%%%%%%%%%
%\bibliography{references_v3} 

\begin{thebibliography}{42}%
	\makeatletter
	\providecommand \@ifxundefined [1]{%
		\@ifx{#1\undefined}
	}%
	\providecommand \@ifnum [1]{%
		\ifnum #1\expandafter \@firstoftwo
		\else \expandafter \@secondoftwo
		\fi
	}%
	\providecommand \@ifx [1]{%
		\ifx #1\expandafter \@firstoftwo
		\else \expandafter \@secondoftwo
		\fi
	}%
	\providecommand \natexlab [1]{#1}%
	\providecommand \enquote  [1]{``#1''}%
	\providecommand \bibnamefont  [1]{#1}%
	\providecommand \bibfnamefont [1]{#1}%
	\providecommand \citenamefont [1]{#1}%
	\providecommand \href@noop [0]{\@secondoftwo}%
	\providecommand \href [0]{\begingroup \@sanitize@url \@href}%
	\providecommand \@href[1]{\@@startlink{#1}\@@href}%
	\providecommand \@@href[1]{\endgroup#1\@@endlink}%
	\providecommand \@sanitize@url [0]{\catcode `\\12\catcode `\$12\catcode
		`\&12\catcode `\#12\catcode `\^12\catcode `\_12\catcode `\%12\relax}%
	\providecommand \@@startlink[1]{}%
	\providecommand \@@endlink[0]{}%
	\providecommand \url  [0]{\begingroup\@sanitize@url \@url }%
	\providecommand \@url [1]{\endgroup\@href {#1}{\urlprefix }}%
	\providecommand \urlprefix  [0]{URL }%
	\providecommand \Eprint [0]{\href }%
	\providecommand \doibase [0]{http://dx.doi.org/}%
	\providecommand \selectlanguage [0]{\@gobble}%
	\providecommand \bibinfo  [0]{\@secondoftwo}%
	\providecommand \bibfield  [0]{\@secondoftwo}%
	\providecommand \translation [1]{[#1]}%
	\providecommand \BibitemOpen [0]{}%
	\providecommand \bibitemStop [0]{}%
	\providecommand \bibitemNoStop [0]{.\EOS\space}%
	\providecommand \EOS [0]{\spacefactor3000\relax}%
	\providecommand \BibitemShut  [1]{\csname bibitem#1\endcsname}%
	\let\auto@bib@innerbib\@empty
	%</preamble>
	\bibitem [{\citenamefont {Hoy}(2018)}]{hoy2018alexa}%
	\BibitemOpen
	\bibfield  {author} {\bibinfo {author} {\bibfnamefont {M.~B.}\ \bibnamefont
			{Hoy}},\ }\href@noop {} {\bibfield  {journal} {\bibinfo  {journal} {Med. Ref.
				Serv. Q.}\ }\textbf {\bibinfo {volume} {37}},\ \bibinfo {pages} {81}
		(\bibinfo {year} {2018})}\BibitemShut {NoStop}%
	\bibitem [{\citenamefont {Raucci}\ \emph {et~al.}(2021)\citenamefont {Raucci},
		\citenamefont {Valentini}, \citenamefont {Pieri}, \citenamefont {Weir},
		\citenamefont {Seritan},\ and\ \citenamefont
		{Mart{\'\i}nez}}]{raucci2021voice}%
	\BibitemOpen
	\bibfield  {author} {\bibinfo {author} {\bibfnamefont {U.}~\bibnamefont
			{Raucci}}, \bibinfo {author} {\bibfnamefont {A.}~\bibnamefont {Valentini}},
		\bibinfo {author} {\bibfnamefont {E.}~\bibnamefont {Pieri}}, \bibinfo
		{author} {\bibfnamefont {H.}~\bibnamefont {Weir}}, \bibinfo {author}
		{\bibfnamefont {S.}~\bibnamefont {Seritan}}, \ and\ \bibinfo {author}
		{\bibfnamefont {T.~J.}\ \bibnamefont {Mart{\'\i}nez}},\ }\href@noop {}
	{\bibfield  {journal} {\bibinfo  {journal} {Nat. Comput. Sci.}\ }\textbf
		{\bibinfo {volume} {1}},\ \bibinfo {pages} {42} (\bibinfo {year}
		{2021})}\BibitemShut {NoStop}%
	\bibitem [{\citenamefont {Wei}(2019)}]{wei2019protein}%
	\BibitemOpen
	\bibfield  {author} {\bibinfo {author} {\bibfnamefont {G.-W.}\ \bibnamefont
			{Wei}},\ }\href@noop {} {\bibfield  {journal} {\bibinfo  {journal} {Nat.
				Mach. Intell.}\ }\textbf {\bibinfo {volume} {1}},\ \bibinfo {pages} {336}
		(\bibinfo {year} {2019})}\BibitemShut {NoStop}%
	\bibitem [{\citenamefont {Mccarthy}\ \emph {et~al.}(2004)\citenamefont
		{Mccarthy}, \citenamefont {Marx}, \citenamefont {Hoffman}, \citenamefont
		{Gee}, \citenamefont {O'neil}, \citenamefont {Ujwal},\ and\ \citenamefont
		{Hotchkiss}}]{mccarthy2004applications}%
	\BibitemOpen
	\bibfield  {author} {\bibinfo {author} {\bibfnamefont {J.~F.}\ \bibnamefont
			{Mccarthy}}, \bibinfo {author} {\bibfnamefont {K.~A.}\ \bibnamefont {Marx}},
		\bibinfo {author} {\bibfnamefont {P.~E.}\ \bibnamefont {Hoffman}}, \bibinfo
		{author} {\bibfnamefont {A.~G.}\ \bibnamefont {Gee}}, \bibinfo {author}
		{\bibfnamefont {P.}~\bibnamefont {O'neil}}, \bibinfo {author} {\bibfnamefont
			{M.~L.}\ \bibnamefont {Ujwal}}, \ and\ \bibinfo {author} {\bibfnamefont
			{J.}~\bibnamefont {Hotchkiss}},\ }\href@noop {} {\bibfield  {journal}
		{\bibinfo  {journal} {Ann. N. Y. Acad. Sci.}\ }\textbf {\bibinfo {volume}
			{1020}},\ \bibinfo {pages} {239} (\bibinfo {year} {2004})}\BibitemShut
	{NoStop}%
	\bibitem [{\citenamefont {Agarap}(2018)}]{agarap2018breast}%
	\BibitemOpen
	\bibfield  {author} {\bibinfo {author} {\bibfnamefont {A.~F.~M.}\
			\bibnamefont {Agarap}},\ }in\ \href@noop {} {\emph {\bibinfo {booktitle}
			{Proceedings of the 2nd international conference on machine learning and soft
				computing}}}\ (\bibinfo {year} {2018})\BibitemShut {NoStop}%
	\bibitem [{\citenamefont {Saba}(2020)}]{saba2020recent}%
	\BibitemOpen
	\bibfield  {author} {\bibinfo {author} {\bibfnamefont {T.}~\bibnamefont
			{Saba}},\ }\href@noop {} {\bibfield  {journal} {\bibinfo  {journal} {Journal
				of Infection and Public Health}\ }\textbf {\bibinfo {volume} {13}},\ \bibinfo
		{pages} {1274} (\bibinfo {year} {2020})}\BibitemShut {NoStop}%
	\bibitem [{\citenamefont {Bhavna}\ and\ \citenamefont
		{Sonawane}(2023)}]{bhavna:NNridges}%
	\BibitemOpen
	\bibfield  {author} {\bibinfo {author} {\bibfnamefont {R.}~\bibnamefont
			{Bhavna}}\ and\ \bibinfo {author} {\bibfnamefont {M.}~\bibnamefont
			{Sonawane}},\ }\href@noop {} {\bibfield  {journal} {\bibinfo  {journal} {npj
				Systems Biology and Applications}\ }\textbf {\bibinfo {volume} {9}},\
		\bibinfo {pages} {21} (\bibinfo {year} {2023})}\BibitemShut {NoStop}%
	\bibitem [{\citenamefont {Wittek}(2014)}]{wittek2014quantum}%
	\BibitemOpen
	\bibfield  {author} {\bibinfo {author} {\bibfnamefont {P.}~\bibnamefont
			{Wittek}},\ }\href@noop {} {\emph {\bibinfo {title} {Quantum machine
				learning: what quantum computing means to data mining}}}\ (\bibinfo
	{publisher} {Academic Press},\ \bibinfo {year} {2014})\BibitemShut {NoStop}%
	\bibitem [{\citenamefont {Banchi}\ \emph {et~al.}(2021)\citenamefont {Banchi},
		\citenamefont {Pereira},\ and\ \citenamefont
		{Pirandola}}]{banchi2021generalization}%
	\BibitemOpen
	\bibfield  {author} {\bibinfo {author} {\bibfnamefont {L.}~\bibnamefont
			{Banchi}}, \bibinfo {author} {\bibfnamefont {J.}~\bibnamefont {Pereira}}, \
		and\ \bibinfo {author} {\bibfnamefont {S.}~\bibnamefont {Pirandola}},\
	}\href@noop {} {\bibfield  {journal} {\bibinfo  {journal} {PRX Quantum}\
		}\textbf {\bibinfo {volume} {2}},\ \bibinfo {pages} {040321} (\bibinfo {year}
		{2021})}\BibitemShut {NoStop}%
	\bibitem [{\citenamefont {Ceriotti}\ \emph {et~al.}(2021)\citenamefont
		{Ceriotti}, \citenamefont {Clementi},\ and\ \citenamefont {Anatole~von
			Lilienfeld}}]{ceriotti2021introduction}%
	\BibitemOpen
	\bibfield  {author} {\bibinfo {author} {\bibfnamefont {M.}~\bibnamefont
			{Ceriotti}}, \bibinfo {author} {\bibfnamefont {C.}~\bibnamefont {Clementi}},
		\ and\ \bibinfo {author} {\bibfnamefont {O.}~\bibnamefont {Anatole~von
				Lilienfeld}},\ }\href@noop {} {\enquote {\bibinfo {title} {Introduction:
				Machine learning at the atomic scale},}\ } (\bibinfo {year}
	{2021})\BibitemShut {NoStop}%
	\bibitem [{\citenamefont {H{\"a}se}\ \emph {et~al.}(2017)\citenamefont
		{H{\"a}se}, \citenamefont {Kreisbeck},\ and\ \citenamefont
		{Aspuru-Guzik}}]{hase2017machine}%
	\BibitemOpen
	\bibfield  {author} {\bibinfo {author} {\bibfnamefont {F.}~\bibnamefont
			{H{\"a}se}}, \bibinfo {author} {\bibfnamefont {C.}~\bibnamefont {Kreisbeck}},
		\ and\ \bibinfo {author} {\bibfnamefont {A.}~\bibnamefont {Aspuru-Guzik}},\
	}\href@noop {} {\bibfield  {journal} {\bibinfo  {journal} {Chem. Sci.}\
		}\textbf {\bibinfo {volume} {8}},\ \bibinfo {pages} {8419} (\bibinfo {year}
		{2017})}\BibitemShut {NoStop}%
	\bibitem [{\citenamefont {Torlai}\ \emph {et~al.}(2019)\citenamefont {Torlai},
		\citenamefont {Timar}, \citenamefont {Van~Nieuwenburg}, \citenamefont
		{Levine}, \citenamefont {Omran}, \citenamefont {Keesling}, \citenamefont
		{Bernien}, \citenamefont {Greiner}, \citenamefont {Vuleti{\'c}},
		\citenamefont {Lukin} \emph {et~al.}}]{torlai2019integrating}%
	\BibitemOpen
	\bibfield  {author} {\bibinfo {author} {\bibfnamefont {G.}~\bibnamefont
			{Torlai}}, \bibinfo {author} {\bibfnamefont {B.}~\bibnamefont {Timar}},
		\bibinfo {author} {\bibfnamefont {E.~P.}\ \bibnamefont {Van~Nieuwenburg}},
		\bibinfo {author} {\bibfnamefont {H.}~\bibnamefont {Levine}}, \bibinfo
		{author} {\bibfnamefont {A.}~\bibnamefont {Omran}}, \bibinfo {author}
		{\bibfnamefont {A.}~\bibnamefont {Keesling}}, \bibinfo {author}
		{\bibfnamefont {H.}~\bibnamefont {Bernien}}, \bibinfo {author} {\bibfnamefont
			{M.}~\bibnamefont {Greiner}}, \bibinfo {author} {\bibfnamefont
			{V.}~\bibnamefont {Vuleti{\'c}}}, \bibinfo {author} {\bibfnamefont {M.~D.}\
			\bibnamefont {Lukin}},  \emph {et~al.},\ }\href@noop {} {\bibfield  {journal}
		{\bibinfo  {journal} {Phys. Rev. Lett.}\ }\textbf {\bibinfo {volume} {123}},\
		\bibinfo {pages} {230504} (\bibinfo {year} {2019})}\BibitemShut {NoStop}%
	\bibitem [{\citenamefont {Chong}\ \emph {et~al.}(2022)\citenamefont {Chong},
		\citenamefont {Kim}, \citenamefont {Ahn},\ and\ \citenamefont
		{Jeong}}]{chong2022machine}%
	\BibitemOpen
	\bibfield  {author} {\bibinfo {author} {\bibfnamefont {D.~R.}\ \bibnamefont
			{Chong}}, \bibinfo {author} {\bibfnamefont {M.}~\bibnamefont {Kim}}, \bibinfo
		{author} {\bibfnamefont {J.}~\bibnamefont {Ahn}}, \ and\ \bibinfo {author}
		{\bibfnamefont {H.}~\bibnamefont {Jeong}},\ }\href@noop {} {\bibfield
		{journal} {\bibinfo  {journal} {Comput. Phys. Commun.}\ }\textbf {\bibinfo
			{volume} {17}},\ \bibinfo {pages} {1} (\bibinfo {year} {2022})}\BibitemShut
	{NoStop}%
	\bibitem [{\citenamefont {Papi{\v{c}}}\ and\ \citenamefont
		{de~Vega}(2022)}]{papivc2022neural}%
	\BibitemOpen
	\bibfield  {author} {\bibinfo {author} {\bibfnamefont {M.}~\bibnamefont
			{Papi{\v{c}}}}\ and\ \bibinfo {author} {\bibfnamefont {I.}~\bibnamefont
			{de~Vega}},\ }\href@noop {} {\bibfield  {journal} {\bibinfo  {journal} {Phys.
				Rev. A}\ }\textbf {\bibinfo {volume} {105}},\ \bibinfo {pages} {022605}
		(\bibinfo {year} {2022})}\BibitemShut {NoStop}%
	\bibitem [{\citenamefont {Luo}\ \emph {et~al.}(2022)\citenamefont {Luo},
		\citenamefont {Chen}, \citenamefont {Carrasquilla},\ and\ \citenamefont
		{Clark}}]{luo2022autoregressive}%
	\BibitemOpen
	\bibfield  {author} {\bibinfo {author} {\bibfnamefont {D.}~\bibnamefont
			{Luo}}, \bibinfo {author} {\bibfnamefont {Z.}~\bibnamefont {Chen}}, \bibinfo
		{author} {\bibfnamefont {J.}~\bibnamefont {Carrasquilla}}, \ and\ \bibinfo
		{author} {\bibfnamefont {B.~K.}\ \bibnamefont {Clark}},\ }\href@noop {}
	{\bibfield  {journal} {\bibinfo  {journal} {Phys. Rev. Lett.}\ }\textbf
		{\bibinfo {volume} {128}},\ \bibinfo {pages} {090501} (\bibinfo {year}
		{2022})}\BibitemShut {NoStop}%
	\bibitem [{\citenamefont {Bandyopadhyay}\ \emph {et~al.}(2018)\citenamefont
		{Bandyopadhyay}, \citenamefont {Huang}, \citenamefont {Sun},\ and\
		\citenamefont {Zhao}}]{bandyopadhyay2018applications}%
	\BibitemOpen
	\bibfield  {author} {\bibinfo {author} {\bibfnamefont {S.}~\bibnamefont
			{Bandyopadhyay}}, \bibinfo {author} {\bibfnamefont {Z.}~\bibnamefont
			{Huang}}, \bibinfo {author} {\bibfnamefont {K.}~\bibnamefont {Sun}}, \ and\
		\bibinfo {author} {\bibfnamefont {Y.}~\bibnamefont {Zhao}},\ }\href@noop {}
	{\bibfield  {journal} {\bibinfo  {journal} {Chem. Phys.}\ }\textbf {\bibinfo
			{volume} {515}},\ \bibinfo {pages} {272} (\bibinfo {year}
		{2018})}\BibitemShut {NoStop}%
	\bibitem [{\citenamefont {Mukherjee}\ \emph {et~al.}(2024)\citenamefont
		{Mukherjee}, \citenamefont {Schachenmayer}, \citenamefont {Whitlock},\ and\
		\citenamefont {W{\"u}ster}}]{mukherjee:modelbuilding}%
	\BibitemOpen
	\bibfield  {author} {\bibinfo {author} {\bibfnamefont {K.}~\bibnamefont
			{Mukherjee}}, \bibinfo {author} {\bibfnamefont {J.}~\bibnamefont
			{Schachenmayer}}, \bibinfo {author} {\bibfnamefont {S.}~\bibnamefont
			{Whitlock}}, \ and\ \bibinfo {author} {\bibfnamefont {S.}~\bibnamefont
			{W{\"u}ster}},\ }\href@noop {} {\bibfield  {journal} {\bibinfo  {journal}
			{arXiv preprint arXiv:2409.18822}\ } (\bibinfo {year} {2024})}\BibitemShut
	{NoStop}%
	\bibitem [{\citenamefont {Poyatos}\ \emph {et~al.}(1997)\citenamefont
		{Poyatos}, \citenamefont {Cirac},\ and\ \citenamefont
		{Zoller}}]{Poyatos_QPT}%
	\BibitemOpen
	\bibfield  {author} {\bibinfo {author} {\bibfnamefont {J.~F.}\ \bibnamefont
			{Poyatos}}, \bibinfo {author} {\bibfnamefont {J.~I.}\ \bibnamefont {Cirac}},
		\ and\ \bibinfo {author} {\bibfnamefont {P.}~\bibnamefont {Zoller}},\
	}\href@noop {} {\bibfield  {journal} {\bibinfo  {journal} {Phys. Rev. Lett.}\
		}\textbf {\bibinfo {volume} {78}},\ \bibinfo {pages} {390} (\bibinfo {year}
		{1997})}\BibitemShut {NoStop}%
	\bibitem [{\citenamefont {Chuang}\ and\ \citenamefont
		{Nielsen}(1997)}]{Chuang_Nielsen_QPT}%
	\BibitemOpen
	\bibfield  {author} {\bibinfo {author} {\bibfnamefont {I.~L.}\ \bibnamefont
			{Chuang}}\ and\ \bibinfo {author} {\bibfnamefont {M.~A.}\ \bibnamefont
			{Nielsen}},\ }\href@noop {} {\bibfield  {journal} {\bibinfo  {journal} {J.
				Mod. Opt.}\ }\textbf {\bibinfo {volume} {44}},\ \bibinfo {pages} {2455}
		(\bibinfo {year} {1997})}\BibitemShut {NoStop}%
	\bibitem [{\citenamefont {Sch{\"o}nleber}\ \emph {et~al.}(2015)\citenamefont
		{Sch{\"o}nleber}, \citenamefont {Eisfeld}, \citenamefont {Genkin},
		\citenamefont {Whitlock},\ and\ \citenamefont {W{\"u}ster}}]{David:Rydagg}%
	\BibitemOpen
	\bibfield  {author} {\bibinfo {author} {\bibfnamefont {D.~W.}\ \bibnamefont
			{Sch{\"o}nleber}}, \bibinfo {author} {\bibfnamefont {A.}~\bibnamefont
			{Eisfeld}}, \bibinfo {author} {\bibfnamefont {M.}~\bibnamefont {Genkin}},
		\bibinfo {author} {\bibfnamefont {S.}~\bibnamefont {Whitlock}}, \ and\
		\bibinfo {author} {\bibfnamefont {S.}~\bibnamefont {W{\"u}ster}},\
	}\href@noop {} {\bibfield  {journal} {\bibinfo  {journal} {Phys. Rev. Lett.}\
		}\textbf {\bibinfo {volume} {114}},\ \bibinfo {pages} {123005} (\bibinfo
		{year} {2015})}\BibitemShut {NoStop}%
	\bibitem [{\citenamefont {Schempp}\ \emph {et~al.}(2015)\citenamefont
		{Schempp}, \citenamefont {G{\"u}nter}, \citenamefont {W{\"u}ster},
		\citenamefont {Weidem{\"u}ller},\ and\ \citenamefont
		{Whitlock}}]{schempp:spintransport}%
	\BibitemOpen
	\bibfield  {author} {\bibinfo {author} {\bibfnamefont {H.}~\bibnamefont
			{Schempp}}, \bibinfo {author} {\bibfnamefont {G.}~\bibnamefont {G{\"u}nter}},
		\bibinfo {author} {\bibfnamefont {S.}~\bibnamefont {W{\"u}ster}}, \bibinfo
		{author} {\bibfnamefont {M.}~\bibnamefont {Weidem{\"u}ller}}, \ and\ \bibinfo
		{author} {\bibfnamefont {S.}~\bibnamefont {Whitlock}},\ }\href@noop {}
	{\bibfield  {journal} {\bibinfo  {journal} {Phys. Rev. Lett.}\ }\textbf
		{\bibinfo {volume} {115}},\ \bibinfo {pages} {093002} (\bibinfo {year}
		{2015})}\BibitemShut {NoStop}%
	\bibitem [{\citenamefont {Nogrette}\ \emph {et~al.}(2014)\citenamefont
		{Nogrette}, \citenamefont {Labuhn}, \citenamefont {Ravets}, \citenamefont
		{Barredo}, \citenamefont {B{\'e}guin}, \citenamefont {Vernier}, \citenamefont
		{Lahaye},\ and\ \citenamefont {Browaeys}}]{nogrette2014single}%
	\BibitemOpen
	\bibfield  {author} {\bibinfo {author} {\bibfnamefont {F.}~\bibnamefont
			{Nogrette}}, \bibinfo {author} {\bibfnamefont {H.}~\bibnamefont {Labuhn}},
		\bibinfo {author} {\bibfnamefont {S.}~\bibnamefont {Ravets}}, \bibinfo
		{author} {\bibfnamefont {D.}~\bibnamefont {Barredo}}, \bibinfo {author}
		{\bibfnamefont {L.}~\bibnamefont {B{\'e}guin}}, \bibinfo {author}
		{\bibfnamefont {A.}~\bibnamefont {Vernier}}, \bibinfo {author} {\bibfnamefont
			{T.}~\bibnamefont {Lahaye}}, \ and\ \bibinfo {author} {\bibfnamefont
			{A.}~\bibnamefont {Browaeys}},\ }\href@noop {} {\bibfield  {journal}
		{\bibinfo  {journal} {Phys. Rev. X}\ }\textbf {\bibinfo {volume} {4}},\
		\bibinfo {pages} {021034} (\bibinfo {year} {2014})}\BibitemShut {NoStop}%
	\bibitem [{\citenamefont {Barredo}\ \emph {et~al.}(2016)\citenamefont
		{Barredo}, \citenamefont {De~L{\'e}s{\'e}leuc}, \citenamefont {Lienhard},
		\citenamefont {Lahaye},\ and\ \citenamefont {Browaeys}}]{barredo2016atom}%
	\BibitemOpen
	\bibfield  {author} {\bibinfo {author} {\bibfnamefont {D.}~\bibnamefont
			{Barredo}}, \bibinfo {author} {\bibfnamefont {S.}~\bibnamefont
			{De~L{\'e}s{\'e}leuc}}, \bibinfo {author} {\bibfnamefont {V.}~\bibnamefont
			{Lienhard}}, \bibinfo {author} {\bibfnamefont {T.}~\bibnamefont {Lahaye}}, \
		and\ \bibinfo {author} {\bibfnamefont {A.}~\bibnamefont {Browaeys}},\
	}\href@noop {} {\bibfield  {journal} {\bibinfo  {journal} {Science}\ }\textbf
		{\bibinfo {volume} {354}},\ \bibinfo {pages} {1021} (\bibinfo {year}
		{2016})}\BibitemShut {NoStop}%
	\bibitem [{\citenamefont {Endres}\ \emph {et~al.}(2016)\citenamefont {Endres},
		\citenamefont {Bernien}, \citenamefont {Keesling}, \citenamefont {Levine},
		\citenamefont {Anschuetz}, \citenamefont {Krajenbrink}, \citenamefont
		{Senko}, \citenamefont {Vuletic}, \citenamefont {Greiner},\ and\
		\citenamefont {Lukin}}]{endres2016atom}%
	\BibitemOpen
	\bibfield  {author} {\bibinfo {author} {\bibfnamefont {M.}~\bibnamefont
			{Endres}}, \bibinfo {author} {\bibfnamefont {H.}~\bibnamefont {Bernien}},
		\bibinfo {author} {\bibfnamefont {A.}~\bibnamefont {Keesling}}, \bibinfo
		{author} {\bibfnamefont {H.}~\bibnamefont {Levine}}, \bibinfo {author}
		{\bibfnamefont {E.~R.}\ \bibnamefont {Anschuetz}}, \bibinfo {author}
		{\bibfnamefont {A.}~\bibnamefont {Krajenbrink}}, \bibinfo {author}
		{\bibfnamefont {C.}~\bibnamefont {Senko}}, \bibinfo {author} {\bibfnamefont
			{V.}~\bibnamefont {Vuletic}}, \bibinfo {author} {\bibfnamefont
			{M.}~\bibnamefont {Greiner}}, \ and\ \bibinfo {author} {\bibfnamefont
			{M.~D.}\ \bibnamefont {Lukin}},\ }\href@noop {} {\bibfield  {journal}
		{\bibinfo  {journal} {Science}\ }\textbf {\bibinfo {volume} {354}},\ \bibinfo
		{pages} {1024} (\bibinfo {year} {2016})}\BibitemShut {NoStop}%
	\bibitem [{\citenamefont {Barredo}\ \emph {et~al.}(2018)\citenamefont
		{Barredo}, \citenamefont {Lienhard}, \citenamefont {De~Leseleuc},
		\citenamefont {Lahaye},\ and\ \citenamefont
		{Browaeys}}]{barredo2018synthetic}%
	\BibitemOpen
	\bibfield  {author} {\bibinfo {author} {\bibfnamefont {D.}~\bibnamefont
			{Barredo}}, \bibinfo {author} {\bibfnamefont {V.}~\bibnamefont {Lienhard}},
		\bibinfo {author} {\bibfnamefont {S.}~\bibnamefont {De~Leseleuc}}, \bibinfo
		{author} {\bibfnamefont {T.}~\bibnamefont {Lahaye}}, \ and\ \bibinfo {author}
		{\bibfnamefont {A.}~\bibnamefont {Browaeys}},\ }\href@noop {} {\bibfield
		{journal} {\bibinfo  {journal} {Nature}\ }\textbf {\bibinfo {volume} {561}},\
		\bibinfo {pages} {79} (\bibinfo {year} {2018})}\BibitemShut {NoStop}%
	\bibitem [{\citenamefont {Wang}\ \emph {et~al.}(2020)\citenamefont {Wang},
		\citenamefont {Shevate}, \citenamefont {Wintermantel}, \citenamefont
		{Morgado}, \citenamefont {Lochead},\ and\ \citenamefont
		{Whitlock}}]{Wang_Rydberg_array_NPJ}%
	\BibitemOpen
	\bibfield  {author} {\bibinfo {author} {\bibfnamefont {Y.}~\bibnamefont
			{Wang}}, \bibinfo {author} {\bibfnamefont {S.}~\bibnamefont {Shevate}},
		\bibinfo {author} {\bibfnamefont {T.~M.}\ \bibnamefont {Wintermantel}},
		\bibinfo {author} {\bibfnamefont {M.}~\bibnamefont {Morgado}}, \bibinfo
		{author} {\bibfnamefont {G.}~\bibnamefont {Lochead}}, \ and\ \bibinfo
		{author} {\bibfnamefont {S.}~\bibnamefont {Whitlock}},\ }\href@noop {}
	{\bibfield  {journal} {\bibinfo  {journal} {Npj Quantum Inf.}\ }\textbf
		{\bibinfo {volume} {6}},\ \bibinfo {pages} {54} (\bibinfo {year}
		{2020})}\BibitemShut {NoStop}%
	\bibitem [{\citenamefont {De~Andrade}\ \emph {et~al.}(2022)\citenamefont
		{De~Andrade}, \citenamefont {Diaz}, \citenamefont {Navas}, \citenamefont
		{Guha}, \citenamefont {Monta{\~n}o}, \citenamefont {Smith}, \citenamefont
		{Raymer},\ and\ \citenamefont {Towsley}}]{de2022quantum}%
	\BibitemOpen
	\bibfield  {author} {\bibinfo {author} {\bibfnamefont {M.~G.}\ \bibnamefont
			{De~Andrade}}, \bibinfo {author} {\bibfnamefont {J.}~\bibnamefont {Diaz}},
		\bibinfo {author} {\bibfnamefont {J.}~\bibnamefont {Navas}}, \bibinfo
		{author} {\bibfnamefont {S.}~\bibnamefont {Guha}}, \bibinfo {author}
		{\bibfnamefont {I.}~\bibnamefont {Monta{\~n}o}}, \bibinfo {author}
		{\bibfnamefont {B.}~\bibnamefont {Smith}}, \bibinfo {author} {\bibfnamefont
			{M.}~\bibnamefont {Raymer}}, \ and\ \bibinfo {author} {\bibfnamefont
			{D.}~\bibnamefont {Towsley}},\ }in\ \href@noop {} {\emph {\bibinfo
			{booktitle} {2022 IEEE International Conference on Quantum Computing and
				Engineering (QCE)}}}\ (\bibinfo {year} {2022})\BibitemShut {NoStop}%
	\bibitem [{\citenamefont {W{\"u}ster}\ and\ \citenamefont
		{Rost}(2018)}]{wuester:review}%
	\BibitemOpen
	\bibfield  {author} {\bibinfo {author} {\bibfnamefont {S.}~\bibnamefont
			{W{\"u}ster}}\ and\ \bibinfo {author} {\bibfnamefont {J.~M.}\ \bibnamefont
			{Rost}},\ }\href@noop {} {\bibfield  {journal} {\bibinfo  {journal} {J. Phys.
				B}\ }\textbf {\bibinfo {volume} {51}},\ \bibinfo {pages} {032001} (\bibinfo
		{year} {2018})}\BibitemShut {NoStop}%
	\bibitem [{\citenamefont {Zhang}\ \emph {et~al.}(2011)\citenamefont {Zhang},
		\citenamefont {Robicheaux},\ and\ \citenamefont {Saffman}}]{zhang2011magic}%
	\BibitemOpen
	\bibfield  {author} {\bibinfo {author} {\bibfnamefont {S.}~\bibnamefont
			{Zhang}}, \bibinfo {author} {\bibfnamefont {F.}~\bibnamefont {Robicheaux}}, \
		and\ \bibinfo {author} {\bibfnamefont {M.}~\bibnamefont {Saffman}},\
	}\href@noop {} {\bibfield  {journal} {\bibinfo  {journal} {Phys. Rev. A}\
		}\textbf {\bibinfo {volume} {84}},\ \bibinfo {pages} {043408} (\bibinfo
		{year} {2011})}\BibitemShut {NoStop}%
	\bibitem [{\citenamefont {Leonhardt}\ \emph {et~al.}(2016)\citenamefont
		{Leonhardt}, \citenamefont {W{\"u}ster},\ and\ \citenamefont
		{Rost}}]{leonhardt2016orthogonal}%
	\BibitemOpen
	\bibfield  {author} {\bibinfo {author} {\bibfnamefont {K.}~\bibnamefont
			{Leonhardt}}, \bibinfo {author} {\bibfnamefont {S.}~\bibnamefont
			{W{\"u}ster}}, \ and\ \bibinfo {author} {\bibfnamefont {J.~M.}\ \bibnamefont
			{Rost}},\ }\href@noop {} {\bibfield  {journal} {\bibinfo  {journal} {Phys.
				Rev. A}\ }\textbf {\bibinfo {volume} {93}},\ \bibinfo {pages} {022708}
		(\bibinfo {year} {2016})}\BibitemShut {NoStop}%
	\bibitem [{\citenamefont {Ravets}\ \emph {et~al.}(2015)\citenamefont {Ravets},
		\citenamefont {Labuhn}, \citenamefont {Barredo}, \citenamefont {Lahaye},\
		and\ \citenamefont {Browaeys}}]{ravets2015measurement}%
	\BibitemOpen
	\bibfield  {author} {\bibinfo {author} {\bibfnamefont {S.}~\bibnamefont
			{Ravets}}, \bibinfo {author} {\bibfnamefont {H.}~\bibnamefont {Labuhn}},
		\bibinfo {author} {\bibfnamefont {D.}~\bibnamefont {Barredo}}, \bibinfo
		{author} {\bibfnamefont {T.}~\bibnamefont {Lahaye}}, \ and\ \bibinfo {author}
		{\bibfnamefont {A.}~\bibnamefont {Browaeys}},\ }\href@noop {} {\bibfield
		{journal} {\bibinfo  {journal} {Phys. Rev. A}\ }\textbf {\bibinfo {volume}
			{92}},\ \bibinfo {pages} {020701} (\bibinfo {year} {2015})}\BibitemShut
	{NoStop}%
	\bibitem [{\citenamefont {Dennis}\ \emph {et~al.}(2012)\citenamefont {Dennis},
		\citenamefont {Hope},\ and\ \citenamefont {Johnsson}}]{xmds:docu}%
	\BibitemOpen
	\bibfield  {author} {\bibinfo {author} {\bibfnamefont {G.~R.}\ \bibnamefont
			{Dennis}}, \bibinfo {author} {\bibfnamefont {J.~J.}\ \bibnamefont {Hope}}, \
		and\ \bibinfo {author} {\bibfnamefont {M.~T.}\ \bibnamefont {Johnsson}},\
	}\href@noop {} {} (\bibinfo {year} {2012}),\ \bibinfo {note}
	{http://www.xmds.org/}\BibitemShut {NoStop}%
	\bibitem [{\citenamefont {Dennis}\ \emph {et~al.}(2013)\citenamefont {Dennis},
		\citenamefont {Hope},\ and\ \citenamefont {Johnsson}}]{xmds:paper}%
	\BibitemOpen
	\bibfield  {author} {\bibinfo {author} {\bibfnamefont {G.~R.}\ \bibnamefont
			{Dennis}}, \bibinfo {author} {\bibfnamefont {J.~J.}\ \bibnamefont {Hope}}, \
		and\ \bibinfo {author} {\bibfnamefont {M.~T.}\ \bibnamefont {Johnsson}},\
	}\href@noop {} {\bibfield  {journal} {\bibinfo  {journal} {Comp. Phys.
				Comm.}\ }\textbf {\bibinfo {volume} {184}},\ \bibinfo {pages} {201} (\bibinfo
		{year} {2013})}\BibitemShut {NoStop}%
	\bibitem [{\citenamefont {Hearst}\ \emph {et~al.}(1998)\citenamefont {Hearst},
		\citenamefont {Dumais}, \citenamefont {Osuna}, \citenamefont {Platt},\ and\
		\citenamefont {Scholkopf}}]{hearst1998support}%
	\BibitemOpen
	\bibfield  {author} {\bibinfo {author} {\bibfnamefont {M.~A.}\ \bibnamefont
			{Hearst}}, \bibinfo {author} {\bibfnamefont {S.~T.}\ \bibnamefont {Dumais}},
		\bibinfo {author} {\bibfnamefont {E.}~\bibnamefont {Osuna}}, \bibinfo
		{author} {\bibfnamefont {J.}~\bibnamefont {Platt}}, \ and\ \bibinfo {author}
		{\bibfnamefont {B.}~\bibnamefont {Scholkopf}},\ }\href@noop {} {\bibfield
		{journal} {\bibinfo  {journal} {IEEE Intelligent Systems and their
				applications}\ }\textbf {\bibinfo {volume} {13}},\ \bibinfo {pages} {18}
		(\bibinfo {year} {1998})}\BibitemShut {NoStop}%
	\bibitem [{\citenamefont {Belgiu}\ and\ \citenamefont
		{Dr{\u{a}}gu{\c{t}}}(2016)}]{belgiu2016random}%
	\BibitemOpen
	\bibfield  {author} {\bibinfo {author} {\bibfnamefont {M.}~\bibnamefont
			{Belgiu}}\ and\ \bibinfo {author} {\bibfnamefont {L.}~\bibnamefont
			{Dr{\u{a}}gu{\c{t}}}},\ }\href@noop {} {\bibfield  {journal} {\bibinfo
			{journal} {ISPRS journal of photogrammetry and remote sensing}\ }\textbf
		{\bibinfo {volume} {114}},\ \bibinfo {pages} {24} (\bibinfo {year}
		{2016})}\BibitemShut {NoStop}%
	\bibitem [{\citenamefont {Mucherino}\ \emph {et~al.}(2009)\citenamefont
		{Mucherino}, \citenamefont {Papajorgji},\ and\ \citenamefont
		{Pardalos}}]{mucherino2009k}%
	\BibitemOpen
	\bibfield  {author} {\bibinfo {author} {\bibfnamefont {A.}~\bibnamefont
			{Mucherino}}, \bibinfo {author} {\bibfnamefont {P.~J.}\ \bibnamefont
			{Papajorgji}}, \ and\ \bibinfo {author} {\bibfnamefont {P.~M.}\ \bibnamefont
			{Pardalos}},\ }in\ \href@noop {} {\emph {\bibinfo {booktitle} {Data mining in
				agriculture}}}\ (\bibinfo  {publisher} {Springer},\ \bibinfo {year} {2009})\
	pp.\ \bibinfo {pages} {83--106}\BibitemShut {NoStop}%
	\bibitem [{\citenamefont {W\"uster}(2017)}]{wuester:immcrad}%
	\BibitemOpen
	\bibfield  {author} {\bibinfo {author} {\bibfnamefont {S.}~\bibnamefont
			{W\"uster}},\ }\href@noop {} {\bibfield  {journal} {\bibinfo  {journal}
			{Phys. Rev. Lett.}\ }\textbf {\bibinfo {volume} {119}},\ \bibinfo {pages}
		{013001} (\bibinfo {year} {2017})}\BibitemShut {NoStop}%
	\bibitem [{\citenamefont {Genkin}\ \emph {et~al.}(2016)\citenamefont {Genkin},
		\citenamefont {Sch\"onleber}, \citenamefont {W\"uster},\ and\ \citenamefont
		{Eisfeld}}]{genkin:markovswitch}%
	\BibitemOpen
	\bibfield  {author} {\bibinfo {author} {\bibfnamefont {M.}~\bibnamefont
			{Genkin}}, \bibinfo {author} {\bibfnamefont {D.~W.}\ \bibnamefont
			{Sch\"onleber}}, \bibinfo {author} {\bibfnamefont {S.}~\bibnamefont
			{W\"uster}}, \ and\ \bibinfo {author} {\bibfnamefont {A.}~\bibnamefont
			{Eisfeld}},\ }\href@noop {} {\bibfield  {journal} {\bibinfo  {journal} {J.
				Phys. B}\ }\textbf {\bibinfo {volume} {49}},\ \bibinfo {pages} {134001}
		(\bibinfo {year} {2016})}\BibitemShut {NoStop}%
	\bibitem [{\citenamefont {Mukherjee}\ \emph {et~al.}(2022)\citenamefont
		{Mukherjee}, \citenamefont {Poddar},\ and\ \citenamefont
		{W\"uster}}]{Mukherjee_binding_dephasing_PhysRevA}%
	\BibitemOpen
	\bibfield  {author} {\bibinfo {author} {\bibfnamefont {K.}~\bibnamefont
			{Mukherjee}}, \bibinfo {author} {\bibfnamefont {S.}~\bibnamefont {Poddar}}, \
		and\ \bibinfo {author} {\bibfnamefont {S.}~\bibnamefont {W\"uster}},\
	}\href@noop {} {\bibfield  {journal} {\bibinfo  {journal} {Phys. Rev. A}\
		}\textbf {\bibinfo {volume} {105}},\ \bibinfo {pages} {L041102} (\bibinfo
		{year} {2022})}\BibitemShut {NoStop}%
	\bibitem [{\citenamefont {Mukherjee}\ \emph {et~al.}(2020)\citenamefont
		{Mukherjee}, \citenamefont {Goswami}, \citenamefont {Whitlock}, \citenamefont
		{W{\"u}ster},\ and\ \citenamefont {Eisfeld}}]{mukherjee2020two}%
	\BibitemOpen
	\bibfield  {author} {\bibinfo {author} {\bibfnamefont {K.}~\bibnamefont
			{Mukherjee}}, \bibinfo {author} {\bibfnamefont {H.~P.}\ \bibnamefont
			{Goswami}}, \bibinfo {author} {\bibfnamefont {S.}~\bibnamefont {Whitlock}},
		\bibinfo {author} {\bibfnamefont {S.}~\bibnamefont {W{\"u}ster}}, \ and\
		\bibinfo {author} {\bibfnamefont {A.}~\bibnamefont {Eisfeld}},\ }\href@noop
	{} {\bibfield  {journal} {\bibinfo  {journal} {New J. Phys.}\ }\textbf
		{\bibinfo {volume} {22}},\ \bibinfo {pages} {073040} (\bibinfo {year}
		{2020})}\BibitemShut {NoStop}%
	\bibitem [{\citenamefont {Mukherjee}\ and\ \citenamefont
		{W{\"u}ster}(2024)}]{Mukherjee_excitons_polaritons}%
	\BibitemOpen
	\bibfield  {author} {\bibinfo {author} {\bibfnamefont {K.}~\bibnamefont
			{Mukherjee}}\ and\ \bibinfo {author} {\bibfnamefont {S.}~\bibnamefont
			{W{\"u}ster}},\ }\href@noop {} {\bibfield  {journal} {\bibinfo  {journal}
			{Quant. Sci. Tech.}\ }\textbf {\bibinfo {volume} {9}},\ \bibinfo {pages}
		{025009} (\bibinfo {year} {2024})}\BibitemShut {NoStop}%
	\bibitem [{\citenamefont {G{\"u}nter}\ \emph {et~al.}(2013)\citenamefont
		{G{\"u}nter}, \citenamefont {Schempp}, \citenamefont {Robert-de
			Saint-Vincent}, \citenamefont {Gavryusev}, \citenamefont {Helmrich},
		\citenamefont {Hofmann}, \citenamefont {Whitlock},\ and\ \citenamefont
		{Weidem{\"u}ller}}]{guenter:EITexpt}%
	\BibitemOpen
	\bibfield  {author} {\bibinfo {author} {\bibfnamefont {G.}~\bibnamefont
			{G{\"u}nter}}, \bibinfo {author} {\bibfnamefont {H.}~\bibnamefont {Schempp}},
		\bibinfo {author} {\bibfnamefont {M.}~\bibnamefont {Robert-de
				Saint-Vincent}}, \bibinfo {author} {\bibfnamefont {V.}~\bibnamefont
			{Gavryusev}}, \bibinfo {author} {\bibfnamefont {S.}~\bibnamefont {Helmrich}},
		\bibinfo {author} {\bibfnamefont {C.}~\bibnamefont {Hofmann}}, \bibinfo
		{author} {\bibfnamefont {S.}~\bibnamefont {Whitlock}}, \ and\ \bibinfo
		{author} {\bibfnamefont {M.}~\bibnamefont {Weidem{\"u}ller}},\ }\href@noop {}
	{\bibfield  {journal} {\bibinfo  {journal} {Science}\ }\textbf {\bibinfo
			{volume} {342}},\ \bibinfo {pages} {954} (\bibinfo {year}
		{2013})}\BibitemShut {NoStop}%
\end{thebibliography}


%%%%%%%%%%%%%%%%%%%%%%%%%%%%%%%%%%%
\end{document}